\documentclass[10pt, aps,pre,showpacs, onecolumn,amsmath,amssymb,floatfix]{revtex4-1}
\usepackage[dvips]{graphicx}
\usepackage{bm,pstricks,longtable}

\def\bfp{\mathbf{p}}
\def\bfq{\mathbf{q}}
\def\caln{\mathcal{N}}
\def\bfl{\mbox{\boldmath$\lambda$}}
\def\bfph{\mbox{\boldmath$\phi$}}

\begin{document}
\title{Integrable random matrix ensembles }
\author{E. Bogomolny, O. Giraud, and \fbox{C. Schmit}}
\affiliation{Univ. Paris-Sud,  CNRS, LPTMS, UMR8626,  Orsay, F-91405, France}
\date{April 19, 2011}

\begin{abstract}
We propose new classes of random matrix ensembles whose statistical properties are intermediate between statistics of Wigner-Dyson random matrices and Poisson statistics. The construction is based on integrable $N$-body classical systems with a random distribution of momenta and coordinates of the particles. The Lax matrices of these systems yield random matrix ensembles whose joint distribution of eigenvalues can be calculated analytically thanks to integrability of the underlying system. Formulas for spacing distributions and level compressibility are obtained for various instances of such ensembles.
\end{abstract}

\maketitle

\section{Introduction}
The theory of random matrices, introduced by Wigner in the 1950s, has proved to be a very useful tool in many fields of physics, from localisation theory to quantum transport (see e.g. \cite{GuhMueWei98} and references therein). In quantum chaos, a well accepted conjecture states that Wigner-Dyson random matrix ensembles describe statistical properties of spectra of quantum systems whose classical counterpart is chaotic \cite{BogGiaSch84}, while statistics  of integrable systems is best described by Poisson statistics of independent random variables \cite{BerTab77}. The corresponding wave functions are extended in the chaotic case and localised in the integrable case. The choice of the random matrix ensemble suited to describe the statistical behaviour of a system depends on the symmetries of that system. In the usual setting \cite{Meh90}, standard random matrix ensembles consist of matrices $M$ with independent Gaussian random elements whose measure is invariant over conjugation  
\begin{equation}
M\longrightarrow U^{-1}\, M\, U,
\label{invariance}
\end{equation}
where $U$ is an arbitrary matrix belonging to one of the three following groups of matrices: unitary, orthogonal, or symplectic. The unitary group defines the Gaussian Unitary Ensemble (GUE), which is supposed to describe statistical properties of energy levels of chaotic systems without time-reversal invariance. The orthogonal group corresponds to Gaussian Orthogonal Ensemble (GOE), used for time-reversal invariant chaotic systems. The symplectic group gives rise to Gaussian Symplectic Ensemble (GSE), applicable to time-reversal chaotic systems with half-integer spin without rotational symmetry. 

Though many extensions and generalisations of random matrices have been proposed in order to best describe various models \cite{zirnbauer}, the existence of a large invariance group as in (\ref{invariance}) remains their characteristic feature. Without such invariance it is very difficult to connect analytically simple properties of matrix elements with complex properties of matrix eigenvalues. For all random matrix ensembles with invariance group it is possible to integrate over unnecessary variables in order to get explicitly the joint distribution of eigenvalues under the form
\begin{equation}
P(\lambda_1,\ldots,\lambda_N)\sim \prod_{i<j}|\lambda_j-\lambda_i|^{\beta}\mathrm{e}^{-\sum_{k}V(\lambda_k)},
\label{joint}
\end{equation}
with $V(x)$ a system-dependent potential and $\beta$ a parameter. For the Gaussian ensembles the potential is quadratic and the parameter $\beta$ is equal to 1 for GOE, 2 for GUE, and 4 for GSE. All correlation functions for invariant ensembles can be calculated analytically \cite{Meh90}. However the resulting formulas are cumbersome. For the nearest-neighbour distribution $P(s)$, instead of the exact expression one often uses  a simple surmise proposed by Wigner. This surmise has correct functional dependence at small and large argument and takes the form
\begin{equation}
P(s)=as^{\beta}\mathrm{e}^{- b s^2}
\label{wigner}
\end{equation} 
with constants $a$ and $b$  determined from the normalisation conditions
\begin{equation}
\int_0^{\infty}P(s)\mathrm{d}s=\int_0^{\infty}s P(s)\mathrm{d}s=1\ .
\label{normalization}
\end{equation}
The Wigner-type surmise for the probability $P(n,s)$ that between two eigenvalues separated by $s$ there exist exactly $n-1$ other levels (with $P(1,s)\equiv P(s)$) is \cite{abul}
\begin{equation}
P(n,s)=a_ns^{d_n}\mathrm{e}^{-b_n s^2},\qquad d_n=n-1+\frac{1}{2}n(n+1)\beta
\label{wigner_n}
\end{equation} 
and $a_n$, $b_n$ are fixed by the normalisations 
\begin{equation}
\int_0^{\infty}P(n,s)\mathrm{d}s=1,\qquad \int_0^{\infty}sP(n,s)\mathrm{d}s=n\ .
\label{normalization_n}
\end{equation}

While for chaotic systems it is possible to argue that eigenstates may statistically be invariant under rotations, this is not the case for more general models. In order to describe statistical properties of such systems one has to consider non-invariant ensembles of random matrices. One of the most investigated examples is the three-dimensional Anderson model \cite{anderson}, with on-site disorder and nearest-neighbour coupling. Depending on the strength of the disorder, it can display metallic behaviour well described by standard random matrix ensembles, or insulator behaviour with Poisson-like spectrum. However, at the metal-insulator transition, spectral statistics are of an intermediate type and are not described by invariant ensembles~\cite{mit}. Similar behaviours have been observed in pseudo-integrable billiards \cite{girdif}, quantum maps corresponding to diffractive classical maps \cite{giraud}, or quantum Hall transitions \cite{huckenstein}. Models have been proposed to describe such intermediate statistics \cite{Gerland}, and random matrix ensembles which possess similar features have been constructed, e.g.~power-law random banded matrix ensembles \cite{PRBM,kravtsov}. 

The main purpose of this paper is to construct random matrix ensembles which are not invariant over rotations of eigenstates, but whose joint distributions of eigenvalues can nevertheless be calculated analytically. A short version of the paper has been published in \cite{prl}. All of these ensembles have intermediate statistics, and for certain of them spectral correlation functions, e.g. the nearest-neighbour distribution, are obtained explicitly. Eigenfunctions of these ensembles are neither localised (as for integrable systems) nor extended (as for chaotic models) but have fractal properties \cite{fractal}. 

Random matrices of the proposed critical ensembles are constructed from the Lax matrices of classical integrable models.  
These models are systems of $N$ classical particles  labelled by an index $i$, $1\leq i \leq N$ in a one-dimensional space. Each particle $i$ is characterised by its position in space $q_i$ and its momentum $p_i$. The dynamics of the particles is entirely described by the Hamiltonian $H(\bfp,\bfq)$, where $\bfp=(p_1,\ldots,p_N)$ and $\bfq=(q_1,\ldots,q_N)$. The characteristic property of these models is the existence of a pair of $N\times N$ matrices $L$ and $M$, called the Lax pair of the system \cite{lax}, such that the equations of motion (the Hamilton equations, derived from the system Hamiltonian) are equivalent to
\begin{equation}
\frac{\partial L}{\partial t}=M\,L-L\,M.
\label{dot_L}
\end{equation}
The Lax matrix $L$ is a matrix depending on momenta $\bfp$ and coordinates $\bfq$. We propose to consider these Lax matrices as random matrices with a certain 'natural' measure of random variables  $p_j$ and $q_j$  
\begin{equation}
\mathrm{d}L=P(\bfp\, ,\bfq\, )\, \mathrm{d}^N\bfp\,\mathrm{d}^N\bfq\ .
\label{measure_pq}
\end{equation} 
The explicit form of this measure depends on the system and will be discussed below. We do not impose any dynamics on variables $\bfp$ and $\bfq$.  The only information we use from the
integrability of the underlying classical system is the existence and
explicit form of action-angle variables $I_{\alpha}(\bfp\, ,\bfq\, )$ and  $\phi_{\alpha}(\bfp\, ,\bfq\, )$. In particular, it is well known that the transformation from momenta and coordinates to action-angle variables is canonical, so that
\begin{equation}
\prod_j\mathrm{d}p_j\,\mathrm{d}q_j=\prod_{\alpha}\mathrm{d}I_{\alpha}\,\mathrm{d}\phi_{\alpha}\ .
\end{equation}
Direct proof that the transformation is canonical is difficult in general, and implicit methods have been used to establish it for specific systems \cite{I}-\cite{III}. In the models we consider here, action variables turn out to be the eigenvalues $\lambda_{\alpha}$ of the Lax matrix, or a simple function of them. The canonical change of variables from momenta and coordinates to action-angle variables in (\ref{measure_pq}) leads to a formal relation
\begin{equation}
\mathrm{d}L={\cal P}(\bfl\, ,\bfph\, )\, \mathrm{d}^N\bfl\,\mathrm{d}^N\bfph\ ,
\label{measure_lphi}
\end{equation}
where ${\cal P}(\bfl\, ,\bfph\, )\equiv  P(\bfp(\bfl \, ,\bfph\, )\, ,\bfq(\bfl \, ,\bfph\, ))$. The exact joint distribution of eigenvalues is then obtained by integration over angle variables, which can easily be performed in all cases considered, and yields
\begin{equation}
P(\bfl)=\int {\cal P}(\bfl\, ,\bfph\, )\mathrm{d}^N\bfph\ . 
\end{equation} 
This scheme is general and can be adapted to many different models. 

In this paper we consider  in detail four typical models of $N$-particle classical integrable systems. The three first, labelled CM$_r$, CM$_h$, and CM$_t$, correspond to the rational, hyperbolic, and trigonometric Calogero-Moser models \cite{calogero,moser}. The fourth model, labelled RS, is a trigonometric variant of the Ruijsenaars-Schneider model \cite{rs}. 

The Calogero-Moser models are defined by the Hamiltonian
\begin{equation}
H(\bfp\, ,\bfq\,)=\frac{1}{2}\sum_j p_j^2+g^2\sum_{j<k}v(q_j-q_k),
\label{H_calogero}
\end{equation}
where $v(\xi)$ is a potential depending on the distance between particles and $g$ is a constant \cite{perelomov}. For the models considered here it has the form $v(\xi)=x^2(\xi)$, where 
\begin{equation}
x(\xi)=\left \{ \begin{array}{ll}\dfrac{1}{\xi}\qquad& \mathrm{model\ CM_r} \\ 
\dfrac{\mu/2}{\sinh (\mu \xi/2)}\qquad &\mathrm{model\ CM_h}\\ 
\dfrac{\mu/2}{\sin (\mu \xi/2 )}\qquad& \mathrm{model\ CM_t} \end{array}\right. \;.
\label{choice_x}
\end{equation}
The Hamiltonian of our fourth model, the trigonometric Ruijsenaars-Schneider model, is \cite{rs} 
\begin{equation}
H(\bfp\, ,\bfq\,)=\sum_{j=1}^N\cos( p_j)
\prod_{k\neq j}\left (1-\frac{\sin^2 \big [  \mu g/2\big ]}{\sin^2\big [\mu (q_j-q_k)/2 \big ]}\right )^{1/2} \hspace{1cm} \mathrm{model\ RS}\; .
\label{hamiltonian_rs}
\end{equation}
The plan of the paper is the following. Sections \ref{model_GM_R}, \ref{model_CM_H}, and \ref{model_CM_T} are devoted to the construction of critical ensembles related respectively with the rational, hyperbolic, and trigonometric Calogero-Moser models. In each of these sections we briefly present the construction of the action-angle variables and choose a 'natural' measure of random momenta and coordinates which allows an easy change of variables as in (\ref{measure_lphi}).  We then give  explicit formulas for the joint distribution of eigenvalues for the resulting critical ensembles of Lax matrices. In section~\ref{model_RS} this scheme is applied to the Ruijsenaars-Schneider model. For this model  the joint distribution of eigenvalues takes a form which makes it suitable for the application of the transfer operator formalism. This approach is detailed in section~\ref{transfer_op}, and in section~\ref{correlation_functions} it is applied to the analytic calculation of nearest-neighbour distributions for the RS model. The spectral compressibility for this model is obtained in section~\ref{compressibility}.\\

For clarity we state below the principal results for the four models considered in this paper.

\subsection*{CM$_r$ ensemble}

The CM$_r$ ensemble is defined as the ensemble of $N\times N$ Hermitian matrices of the form
\begin{equation}
L_{kr}=p_r\delta_{kr}+\mathrm{i}g\dfrac{1-\delta_{kr}}{q_k-q_r}\ ,
\end{equation}
with $g$ a real constant. Positions $\bfq$ and momenta $\bfp$ are random variables distributed according to the density 
\begin{equation}
P(\bfp\, ,\bfq\, )\sim \exp \left [-A\Big (\sum_j p_j^2+g^2\sum_{j\neq k}\frac{1}{(q_j-q_k)^2}\Big )-B\sum_jq_j^2\right ]\ ,
\end{equation}
with $A$ and $B$ arbitrary positive constants. The joint distribution  of eigenvalues for this ensemble is then given by
\begin{equation}
P(\bfl)\sim \exp \left [-A\sum_{\alpha} \lambda_{\alpha}^2 -
g^2 B \sum_{\alpha\neq \beta}\frac{1}{(\lambda_{\alpha}-\lambda_{\beta})^2}  \right ]\ .
\end{equation}
A characteristic property of this ensemble is the exponentially strong level repulsion: the nearest-neighbour spacing distribution $P(s)$ is characterised by
\begin{equation}
\ln P(s)\underset{s\to 0}{\sim} -\frac{b}{s^2}+\mathcal{O}(1).
\label{p_CM_R}
\end{equation}
We propose the following Wigner-type surmise for the next-to-nearest-neighbour spacing distributions $P(n,s)$, depending on four parameters:
\begin{equation}
P(n,s)=as^d\exp \left (-\frac{b}{s^2} -c s\right ).
\end{equation}
It contains two fitting constants depending on $n$. The other two are fixed by the normalisation \eqref{normalization_n}.

\subsection*{CM$_h$ ensemble}

The CM$_h$ ensemble is defined as the ensemble of $N\times N$ Hermitian matrices of the form
\begin{equation}
L_{kr}=p_r\delta_{kr}+\mathrm{i}g\dfrac{\mu(1-\delta_{kr})}{2\sinh \big [\mu (q_k-q_r)/2\big ]}
\label{Lax_CM_H}
\end{equation}
with $g$ and $\mu$ real constants, and $\bfq$ and $\bfp$ distributed according to the density 
\begin{equation}
P(\bfp\, ,\bfq\, ) \sim \exp \left[-A\Big (\sum_j p_j^2+g^2\sum_{j\neq k}
\frac{\mu^2}{4\sinh^2 \big [\mu (q_j-q_k)/2\big ] } \Big )-B\sum_j \cosh \mu q_j\right ]\ .
\end{equation}
The exact joint distribution for this model is
\begin{equation}
P(\bfl)\sim \exp\left(-A\sum_{\alpha} \lambda_{\alpha}^2\right)\prod_{\alpha}K_0 \left ( B\prod_{\beta \neq \alpha }
\Big |1+\frac{\mathrm{i}g \mu }{\lambda_{\alpha}-\lambda_{\beta}} \Big | \right )
\end{equation}
where $K_0(x)$ is the modified Bessel function of the second kind. The nearest-neighbour spacing distribution has an exponential asymptotic similar to (\ref{p_CM_R}) but with $1/s$  leading term instead of $1/s^2$, namely
\begin{equation}
\ln P(s)\underset{s\to 0}{\sim} -\frac{b}{s}+\mathcal{O}(\ln s)\, .
\label{p_CM_H}
\end{equation}
The Wigner-type surmise for CM$_h$ is  
\begin{equation}
P(n,s)=as^d\exp \left (-\frac{b}{s} -c s\right ).
\end{equation}

\subsection*{CM$_t$ ensemble}

Matrices from this ensemble correspond to a situation where $\mu$ in Eq.~(\ref{Lax_CM_H}) is allowed to take pure imaginary values. They are of the form
\begin{equation}
L_{kr}=p_r\delta_{kr}+\mathrm{i}g\dfrac{\mu(1-\delta_{kr})}{2\sin \big [\mu(q_k-q_r)/2\big ]}
\label{Lax_CM_T}
\end{equation}
with $g$ and $\mu$ real constants, and $\bfq$ and $\bfp$ distributed according to the density 
\begin{equation}
P(\bfp\, ,\bfq\, ) \sim \exp \left [-A\Big (\sum_j p_j^2+g^2\sum_{j\neq k}\frac{\mu^2}{4\sin^2 \big [\mu(q_j-q_k)/2\big ] } \Big ) \right ]
\end{equation} 
with the restrictions that all $q_j$ are between $0$ and $2\pi/\mu$. The exact joint distribution of eigenvalues for this ensemble is
\begin{equation}
P(\bfl)\sim \exp \left(-A\sum_{\alpha} \lambda_{\alpha}^2\right)\chi(\bfl)\ ,
\end{equation}
where the function $\chi(\bfl)$ is equal to 1 if $\lambda_1<\lambda_2<\ldots<\lambda_N$ and $\lambda_{\alpha+1}-\lambda_{\alpha}>\mu g$ for all $\alpha$. The nearest-neighbour spacing distribution is given by a shifted Poisson distribution of the form
\begin{equation}
P(s)=\left \{\begin{array}{ll}0,&  0<s<b \\
\dfrac{1}{1-b}\exp \left (-\dfrac{s-b}{1-b}\right ),&s>b \end{array}\right.
\label{shifted_Poisson}
\end{equation} 
with $b$ some fitting constant.

\subsection*{RS ensemble}
The RS ensemble is defined as the ensemble of $N\times N$ matrices of the form~\cite{rs}
\begin{equation}
L_{kr}=\mathrm{e}^{\mathrm{i}\sigma p_k/2}\tilde{W}_k^{1/2}
\dfrac{\sin \big [\mu g\sigma/2\big ] }{\sin \big [\mu (q_k-q_r+g\sigma)/2\big ]}
\tilde{V}_r^{1/2}\mathrm{e}^{\mathrm{i}\sigma p_r/2}
\label{Lax_PS}
\end{equation}
with
\begin{equation}
\tilde{V}_k=\prod_{j\neq k}\frac{\sin \big [\mu (q_k-q_j-g\sigma)/2\big ]}{\sin \big [\mu (q_k-q_j)/2\big ]}\ ,\qquad
\tilde{W}_k=\prod_{j\neq k}\frac{\sin \big [\mu (q_k-q_j+g\sigma)/2\big ]}{\sin \big [\mu (q_k-q_j)/2\big ]}\ .
\end{equation}
Let  $\tilde{\Omega}$ be the set of $\bfq$ such that for all $k$ the sign of both $\tilde{V}_k$ and $\tilde{W}_k$ is the same as the sign of $\sin (N \mu g\sigma/2)/\sin (\mu g\sigma/2)$. The matrix $L$ is unitary if and only if $\bfq\in\tilde{\Omega}$. The variables $\bfq$ and $\bfp$ are chosen to be distributed according to the uniform density in the region where $L$ is unitary. That is, we choose momentum variables $p_j$ independent and uniformly distributed between $0$ and $2\pi/\sigma$ and coordinate variables $\bfq$ uniformly distributed over $\tilde{\Omega}$. In this case eigenvalues of the Lax matrices (\ref{Lax_PS}) are also  uniformly distributed over $\tilde{\Omega}$. Choosing $\mu=2\pi /N$, $\sigma=1$ and $g=a$, we compute correlation functions of eigenvalues of matrix (\ref{Lax_PS}) for fixed $a$ and $N\to \infty$. The results strongly depend on the integer part of $a$. For $0<a<1$ the nearest-neighbour spacing distribution is similar to (\ref{shifted_Poisson}) with constant $b$ now equal to $a$. For $1<a<2$ the nearest-neighbour distribution takes the form
\begin{equation}
P(s)=\left \{\begin{array}{cl}A^2\sinh^2(\rho s)&\;\;\mathrm{when}\;\;1<g<4/3\\
\frac{81}{64}s^2&\;\;\mathrm{when}\;\;g=4/3\\
A^2\sin^2(\rho s)&\;\;\mathrm{when}\;\;4/3<g<2
\end{array}\right. .
\end{equation}
Constants $A$ and $\rho$ are  determined from the normalisation conditions \eqref{normalization}. Other correlation functions are also obtained in section~\ref{correlation_functions}.

\subsection*{Numerical implementation}
\label{numericalimplementation}
The results presented above are quite robust with respect to alterations in the distribution of $\bfq$ and $\bfp$. In all models considered we chose (as explained above) a distribution of coordinates such that the $q_j$ are confined to a finite interval while having a strong repulsion between each other. As may be expected physically (though we do not have a rigorous proof for this), numerical evidence shows that, if we keep these two characteristic features, spectral properties for $N\to\infty$ depend only weakly on the precise choice for the distribution of $\bfq$ and $\bfp$. 

From these considerations it is thus natural to use, rather than the exact complicated distribution of $\bfq$, the picket-fence configuration when all coordinates are just fixed  and equally spaced. As all definitions of our ensembles involve only differences  $q_j-q_k$ multiplied by a parameter ($\mu$ or $g$, depending on the model), we can without loss of generality choose to take $q_j=j$, $j=1,\ldots N$.

For numerical implementation we chose $q_j=j$, and $p_j$ as independent Gaussian variables with zero mean and with variance equal $1$ (CM ensembles) or independent variables uniformly distributed between $0$ and $2\pi$ (RS ensemble). For concreteness, we fixed $\mu=4\pi/N$ for CM$_h$ and CM$_t$ and $\mu=2\pi/N$ for RS. For such a choice, the $N\times N$ Lax matrices take the form   
\begin{equation}
\begin{array}{ll}
L_{kr}=p_k\delta_{kr}+\mathrm{i}g\dfrac{1-\delta_{kr}}{k-r}, & \mathrm{model\ CM_r}\\
L_{kr}=p_k\delta_{kr}+\mathrm{i}g\dfrac{2\pi(1-\delta_{kr})}{N\sinh \big [2\pi(k-r)/N\big] }, & \mathrm{model\ CM_h}\\
L_{kr}=p_k\delta_{kr}+\mathrm{i}g\dfrac{2\pi(1-\delta_{kr})}{N\sin\big [2\pi(k-r)/N\big] }, & \mathrm{model\ CM_t}\\
L_{kr}=\mathrm{e}^{\mathrm{i}p_k}\dfrac{1-\mathrm{e}^{2\mathrm{i}\pi g }}{N(1-\mathrm{e}^{2\mathrm{i}\pi(k-r+g)/N})}, & \mathrm{model\ RS}\;\; \end{array} \ . 
\label{simplified}
\end{equation}   
For CM$_t$ matrices with even $N$, to avoid the singularity we changed $N\to N+1$ in the above formula. As the figures in the next sections show, despite this particular choice for the distribution of $\bfq$ and $\bfp$, the agreement between the computed spectral statistics and analytical formulas is remarkable.

%===========================================================================================  
\section{Rational Calogero-Moser model}\label{model_GM_R}
The first model we consider is the rational Calogero-Moser model CM$_r$~\cite{perelomov}, characterised by the Lax matrix
\begin{equation}
L_{kr}=p_r\delta_{kr}+\mathrm{i}g\,\frac{1-\delta_{kr}}{q_k-q_r}\ .
\label{Inr}
\end{equation}
It depends on a real constant $g$ and on a set of $2N$ random variables $p_k$ and $q_k$ whose distribution will be specified later on. We are interested in eigenvalues $\lambda_{\alpha}$ and eigenfunctions $u_k(\alpha)$ of this matrix (here and below we will use the Greek letters to label eigenvalues and corresponding eigenfunctions)
\begin{equation}
\label{def_Lu}
\sum_{r=1}^N L_{kr}u_r(\alpha)=\lambda_{\alpha} u_k(\alpha)\ .
\end{equation}
To construct angle-action variables let us define the new quantities
\begin{equation}
\label{defQ_CMR}
Q_{\alpha \beta}=\sum_k u_k^*(\alpha)q_ku_k(\beta). 
\end{equation}
From Eq.~\eqref{Inr} one gets
\begin{equation}
L_{kr}q_r-q_k L_{kr}=-\mathrm{i}g(1-\delta_{kr})\ .
\end{equation}
Multiplying both sides by $u_{k}^{*}(\alpha)u_r(\beta)$ and summing over $k$ and $r$ one gets
\begin{equation}
Q_{\alpha \beta}(\lambda_{\alpha}-\lambda_{\beta})=-\mathrm{i}g(e_{\alpha}^*e_{\beta}-\delta_{\alpha \beta}),
\label{Qmn}
\end{equation}
where 
\begin{equation}
e_{\alpha}=\sum_k u_k(\alpha)\ .
\end{equation}
For $\alpha=\beta$, Eq.~\eqref{Qmn} implies that $|e_{\alpha}|^2=1$, and one can choose the overall phase of the eigenvector $u_k(\alpha)$ in such a way that $e_{\alpha}=1$. Let $\phi_{\alpha}$ be new variables defined by
\begin{equation}
\label{def_phi}
Q_{\alpha\alpha}=\phi_{\alpha}\ .
\end{equation}
Then from Eq.~(\ref{Qmn}) we have 
\begin{equation}
Q_{\alpha \beta}=\phi_{\alpha}\delta_{\alpha \beta}-\mathrm{i}g\frac{1-\delta_{\alpha \beta}}{\lambda_{\alpha}-\lambda_{\beta}}\ .
\label{QInr}
\end{equation} 
The matrix $Q$ can be seen as the dual matrix of $L$, with $\phi_{\alpha}$ playing the role of momenta and $\lambda_{\alpha}$ the role of positions. In \cite{I} it was proved that there is a canonical transformation from position and momentum variables $(q_k, p_k)$ to action and angle variables $(\lambda_{\alpha},\phi_{\alpha})$. Showing that the transformation is canonical is a rather technical mathematical result. However one can easily check that the new variables $\lambda_{\alpha}$ and $\phi_{\alpha}$ verify Hamilton-Jacobi equations (see Appendix \ref{HJ}).

We now consider an ensemble of Hermitian matrices of the form \eqref{Inr} with random variables $p_k$ and $q_k$ drawn according to the measure
\begin{equation}
P(L)dL=\caln  \exp \left [-A\mathrm{Tr} L^2 -B\sum_k q_k^2\right ]\prod_k\mathrm{d}p_k\, \mathrm{d}q_k\ ,
\label{measureInr}
\end{equation}
where $A$ and $B$ are given constants and $\caln$ a normalisation factor. The first term in Eq.~\eqref{measureInr} is the analog of the  usual Gaussian weight of RMT; the second term is a quadratic confinement potential. Since the action-angle transformation is canonical one has
\begin{equation}
\prod_k\mathrm{d}p_k\, \mathrm{d}q_k= \prod_{\alpha}\mathrm{d}\lambda_{\alpha}\mathrm{d}\phi_{\alpha}\ .
\label{canonical}
\end{equation}
From Eq.~\eqref{defQ_CMR}, using orthogonality of eigenvectors one gets
\begin{equation}
\mathrm{Tr}Q^2=\sum_jq_j^2\, .
\end{equation}
Using these relations one can rewrite the distribution (\ref{measureInr}) in action-angle variables $\lambda_{\alpha}$ and $\phi_{\alpha}$ as
\begin{equation}
P(L)dL=\caln
\exp \left [-A \sum_{\alpha} \lambda_{\alpha}^2-B \left (\sum_{\alpha} \phi_{\alpha}^2+g^2\sum_{\alpha\neq \beta}\frac{1}{(\lambda_{\alpha}-\lambda_{\beta})^2}\right ) 
\right ]\prod_{\alpha}\mathrm{d}\lambda_{\alpha}\, \mathrm{d}\phi_{\alpha}\,.
\end{equation}
Integration over the $\phi_{\alpha}$ gives a constant. We thus obtain the joint distribution of eigenvalues for the ensemble of random matrices $L$ with the measure (\ref{measureInr}) as
\begin{equation}
P(\lambda_1,\ldots, \lambda_N)\sim \exp \left [-A \sum_{\alpha} \lambda_{\alpha}^2-B g^2\sum_{\alpha\neq\beta}\frac{1}{(\lambda_{\alpha}-\lambda_{\beta})^2}
\right ]\ . 
\label{distributionInr}
\end{equation}
Note that, similarly as in the standard RMT case, this joint eigenvalue distribution can be interpreted via the Coulomb gas model as the partition function of an ensemble of particles on a line, here with inverse square repulsion. After rescaling $x_k=\lambda_k(Bg^2/A)^{-1/4}$, equilibria positions of the particles at positions $\lambda_{\alpha}$ are given by
\begin{equation}
x_k=2  \sum_{j\neq k}\frac{1}{(x_j-x_k)^3}\,\, , 1\leq k\leq N\ . 
\end{equation}
Such a relation characterises the zeros of Hermite polynomials of degree $N$ (see also Eq.~(10.3) of \cite{perelomov}). It is known from RMT \cite{Meh90} that the distribution of eigenvalues of Gaussian random ensembles has a similar property, which implies that the asymptotic density of eigenvalues is given by Wigner's semi-circle law.

An immediate consequence of the distribution \eqref{distributionInr} is the unusual very strong level repulsion at small distances. For all standard random matrix ensembles the nearest-neighbour distribution $P(s)$  behaves as $s^{\beta}$ at small $s$. By contrast, in our case it follows from (\ref{distributionInr}) that 
\begin{equation}
P(s)\underset{s\to 0}{\sim} a\mathrm{e}^{-b/s^2}.
\end{equation}
As the potential between eigenvalues decreases as the inverse square of the distance between them, the probability of having a gap of size $s$ for large $s$ is exponentially small. We could not calculate exactly correlation functions for the distribution (\ref{distributionInr}). However, combining the two asymptotics above, we build a Wigner-type surmise for the  nearest-neighbour spacing distribution of the form 
\begin{equation}
P(s)=a\mathrm{e}^{-b/s^2-c s}\ ,
\label{surmise}
\end{equation}
where $b$ is a fitting constant, and constants $a$ and $c$ are determined from the normalisation conditions (\ref{normalization}). For the $n$th nearest-neighbour spacing distributions $P(n,s)$, with $n\geq 2$, we conjecture, by analogy with the Wigner surmise~\eqref{wigner_n} for standard random matrices, the form
\begin{equation}
P(n,s)=a s^d \mathrm{e}^{-b/s^2- c s}
\label{surmise2}
\end{equation}
with two fitting constants $b$ and $d$.

\begin{figure}[hbt]
\begin{center}
\includegraphics[width=.65\linewidth]{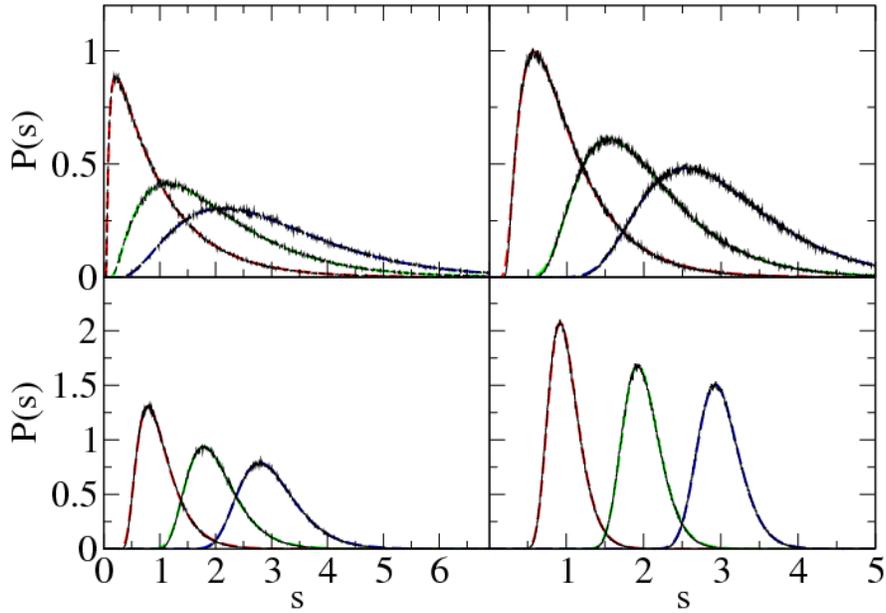}
\end{center}
\caption{(Color online). Nearest-neighbour distributions $P(n,s)$ for the random matrices of model CM$_r$ with $g=0.05$ (top left), $0.25$ (top right), $0.5$ (bottom left), and $1$ (bottom right), averaged over the central quarter of the unfolded spectrum for $8000$ realisations of matrices of size $N=512$. Solid lines are numerical results, dashed lines indicate the fits (\ref{surmise}) for and \eqref{surmise2} for $P(n,s)$ with (in each panel from left to right) $P(s)=P(1,s)$ (red), $P(2,s)$ (green) and $P(3,s)$ (blue). \label{psInr}}
\end{figure}

To assess this conjecture we compare the analytical expressions \eqref{surmise}--\eqref{surmise2} with numerical results, with the choice of parameters detailed in section \ref{numericalimplementation}. Results displayed in Fig.~\ref{psInr} show that the agreement is remarkable.

%===========================================================================
\section{Hyperbolic Calogero-Moser model}\label{model_CM_H}

The Lax matrix for the hyperbolic Calogero-Moser model CM$_h$ reads \cite{perelomov}
\begin{equation}
L_{kr}=p_r\delta_{kr}+\mathrm{i}g(1-\delta_{kr})\dfrac{\mu}{2\sinh(\mu(q_k-q_r)/2)}\ .
\label{IInr}
\end{equation}
Let us define two matrices $Q$ and $R$ by
\begin{equation}
Q_{\alpha \beta}=\sum_k u_k^*(\alpha)\mathrm{e}^{\mu q_k}u_k(\beta)\ ,\qquad
R_{\alpha \beta}=\sum_k u_k^*(\alpha)\mathrm{e}^{-\mu q_k}u_k(\beta)\ ,
\label{Q_IInr}
\end{equation}
and two vectors
\begin{equation}
e_{\alpha}=\sum_k u_k(\alpha)\mathrm{e}^{\mu q_k/2}\ ,\qquad
f_{\alpha}=\sum_k u_k(\alpha)\mathrm{e}^{-\mu q_k/2}\ .
\label{e_alpha}
\end{equation}
From \eqref{IInr} one can get the two equivalent equations
\begin{eqnarray}
\mathrm{e}^{\mu q_k}L_{kr}-L_{kr}\mathrm{e}^{\mu q_r}&=&\mathrm{i}g\mu (1-\delta_{kr})\mathrm{e}^{\mu (q_k+q_r)/2}\\
\mathrm{e}^{-\mu q_r}L_{kr}-L_{kr}\mathrm{e}^{-\mu q_k}&=&\mathrm{i}g\mu (1-\delta_{kr})\mathrm{e}^{-\mu (q_k+q_r)/2}\ .
\label{identity}
\end{eqnarray}
Multiplying both sides by $u_k(\alpha)^* u_r(\beta)$ and summing over all $k$ and $r$ one gets 
\begin{eqnarray}
Q_{\alpha \beta}(\lambda_{\alpha}-\lambda_{\beta})=-\mathrm{i}g\mu (e_{\alpha}^*e_{\beta}-Q_{\alpha \beta})\label{QR1}\\
R_{\alpha \beta}(\lambda_{\alpha}-\lambda_{\beta})=\mathrm{i}g\mu (f_{\alpha}^*f_{\beta}-R_{\alpha \beta})\ ,
\label{QR2}
\end{eqnarray}
which implies that matrices $Q$ and $R$ take the form
\begin{equation}
Q_{\alpha \beta}=e_{\alpha}^*\dfrac{\mathrm{i}g\mu}{\lambda_{\beta}-\lambda_{\alpha}+\mathrm{i}g\mu }e_{\beta}\ ,\qquad
R_{\alpha \beta}=f_{\alpha}^*\dfrac{-\mathrm{i}g\mu}{\lambda_{\beta}-\lambda_{\alpha}-\mathrm{i}g\mu }f_{\beta}\ .
\label{Q_IInr_new}
\end{equation}
By their definition \eqref{Q_IInr}, matrices $Q$ and $R$ are inverse of each other, so that $\sum_{\gamma}Q_{\alpha \gamma}R_{ \gamma \beta}=\delta_{\alpha \beta}$ for all $\alpha,\beta$. For $\alpha=\beta$ this condition implies that
\begin{equation}
-g^2\mu^2 e_{\alpha}^*f_{\alpha}\sum_{\gamma}\frac{e_{\gamma}f_{\gamma}^*}{(\lambda_{\gamma}-\lambda_{\alpha}+\mathrm{i}g\mu)^2}=1
\label{conditionQR1}
\end{equation}
(in particular, it follows that all $e_{\alpha}$ and $f_{\alpha}$ are nonzero). For $\alpha\neq\beta$ one obtains
\begin{equation}
\sum_{\gamma}\frac{e_{\gamma}f_{\gamma}^*}{(\lambda_{\beta}-\lambda_{\alpha}+\mathrm{i}g\mu)(\lambda_{\gamma}-\lambda_{\beta}-\mathrm{i}g\mu ) }=0\ .
\end{equation}
Using the identity 
\begin{equation}
\dfrac{1}{(\lambda_{\beta}-\lambda_{\alpha}+\mathrm{i}g\mu)(\lambda_{\gamma}-\lambda_{\beta}-\mathrm{i}g\mu ) }=
\left [\dfrac{1}{\lambda_{\beta}-\lambda_{\gamma}+\mathrm{i}g\mu}-
\dfrac{1}{\lambda_{\beta}-\lambda_{\alpha}+\mathrm{i}g\mu}\right ]\dfrac{1}{\lambda_{\alpha}-\lambda_{\gamma}}\ ,
\end{equation}
valid for $\alpha\neq \beta$, one concludes that 
\begin{equation}
\sum_{\gamma}\frac{e_{\gamma}f_{\gamma}^*}{\lambda_{\gamma}-\lambda_{\alpha}+\mathrm{i}g\mu}=c\ ,
\label{conditionQR2}
\end{equation}
where $c$ is a certain constant independent on $\alpha$. According to this equation the quantities $b_{\gamma}=e_{\gamma}f_{\gamma}^*/c$ obey a system of linear equations of the form
\begin{equation}
\sum_{\gamma}\frac{b_{\gamma}}{x_{\gamma}-y_{\alpha}}=1\ ,
\label{equation}
\end{equation}
with $x_{\gamma}=\lambda_{\gamma}$ and $y_{\alpha}=\lambda_{\alpha}-\mathrm{i}g\mu$. This equation coincides with Eq.~\eqref{eqn} in Appendix~\ref{app_A}. From \eqref{bm} it follows that 
\begin{equation}
e_{\alpha}f_{\alpha}^*=\mathrm{i}g\mu c V_{\alpha},
\end{equation}
where 
\begin{equation}
V_{\alpha}=\prod_{\beta\neq \alpha}\left(1+\dfrac{\mathrm{i}g\mu}{\lambda_{\alpha}-\lambda_{\beta}}\right)\ ,
\label{V_alphaapp}
\end{equation}
while \eqref{sum_bm} implies that
\begin{equation}
\sum_{\alpha}e_{\alpha}f_{\alpha}^*=\mathrm{i}g\mu c N.
\end{equation}
It readily follows from the definition \eqref{e_alpha} of $e_{\alpha}$ and $f_{\alpha}$ that $\sum_{\alpha}e_{\alpha}f_{\alpha}^*=N$, thus the value of $c$ is fixed by $\mathrm{i}g\mu c=1$. Equation \eqref{conditionQR1} is then fulfilled as a direct consequence of \eqref{sum_bm2}. Finally we have
\begin{equation}
e_{\alpha}f_{\alpha}^*=V_{\alpha}.
\label{efVapp}
\end{equation}
Let $\phi_{\alpha}$ be new variables defined from diagonal elements of  matrix $Q$ by
\begin{equation}
Q_{\alpha \alpha}=|V_{\alpha}|\mathrm{e}^{\mu \phi_{\alpha}}.
\label{Qaa}
\end{equation}
Then from \eqref{Q_IInr_new} and \eqref{efVapp} it follows that
\begin{equation}
R_{\alpha \alpha}=|V_{\alpha}|\mathrm{e}^{-\mu \phi_{\alpha}}.
\label{Raa}
\end{equation} 
In the definition \eqref{e_alpha} of $e_{\alpha}$ it is convenient to choose  the overall phase of the eigenvector $u_k(\alpha)$ in such a way that $e_{\alpha}$ be real. As $Q_{\alpha \alpha}=|e_{\alpha}|^2$ one has
\begin{equation}
e_{\alpha}=|V_{\alpha}|^{1/2}\mathrm{e}^{\mu \phi_{\alpha}/2}.
\label{resultat_e}
\end{equation} 
Using Eq.~\eqref{Q_IInr_new}, the matrix $Q$ can now be expressed in terms of the new variables $\lambda_{\alpha}$ and $\phi_{\alpha}$, as 
\begin{equation}
Q_{\alpha \beta}=|V_{\alpha}|^{1/2}\mathrm{e}^{\mu \phi_{\alpha}/2}
\dfrac{\mathrm{i}g\mu}{\lambda_{\beta}-\lambda_{\alpha}+\mathrm{i}g\mu}\mathrm{e}^{\mu \phi_{\beta}/2}|V_{\beta}|^{1/2}\ .
\label{Qnewvar}
\end{equation}
As in the case of model CM$_r$, the matrix $Q$ can be seen as the dual matrix of $L$. Indeed, $Q$ coincides with the Lax matrix of the rational Ruijsenaars-Schneider  model with coordinates $\lambda_{\alpha}$ and momenta $\phi_{\alpha}$ \cite{I}. In \cite{I} it has been proved that the transformation from position and momentum variables $(q_k,p_k)$ to action-angle variables $(\lambda_{\alpha},\phi_{\alpha})$ is canonical. Again one can check that the new variables $\lambda_{\alpha}$ and $\phi_{\alpha}$ verify Hamilton-Jacobi equations. 

\begin{figure}[hbt]
\begin{center}
\includegraphics[width=.65\linewidth]{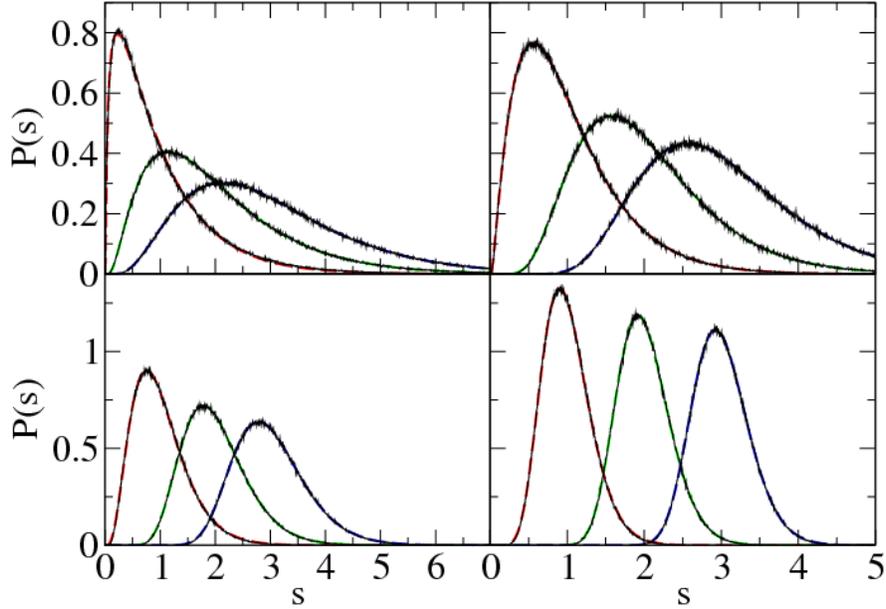}
\end{center}
\caption{(Color online). Nearest-neighbour spacing distributions $P(n,s)$ for the random matrices of model CM$_h$ with $\mu=4\pi/N$ and $g=0.05$ (top left), $0.25$ (top right), $0.5$ (bottom left), and $1$ (bottom right), averaged over the central quarter of the spectrum for $32000$ realisations of matrices of size $N=256$. Solid lines are numerical results, dashed lines indicate the fit \eqref{surmiseIInr} for $P(n,s)$ with (in each panel from left to right) $P(s)=P(1,s)$ (red), $P(2,s)$ (green) and $P(3,s)$ (blue). \label{psIInr}}
\end{figure}

We now consider an ensemble of Hermitian matrices of the form \eqref{IInr} with random variables $p_k$ and $q_k$ drawn according to the measure
\begin{equation}
P(L)dL=\caln  \exp \left [-A\mathrm{Tr} L^2 -B\sum_k \cosh\mu q_k\right ]\prod_k\mathrm{d}p_k\, \mathrm{d}q_k\ .
\label{measureIInr}
\end{equation}
As in the case of model CM$_r$, Eq.~\eqref{measureIInr} contains a standard RMT Gaussian weight and a confinement potential which can be rewritten as $\mathrm{Tr} Q+\mathrm{Tr} R$. Using \eqref{Qaa}, \eqref{Raa}, and  the fact that the transformation is canonical, we get the distribution in terms of the new variables $\lambda_{\alpha}$ and $\phi_{\alpha}$ as
\begin{equation}
P(L)dL=\caln  \exp \left [-A\sum_{\alpha}\lambda_{\alpha}^2 -B\sum_{\alpha}
|V_{\alpha}|\cosh\mu\phi_{\alpha}\right ]\prod_\alpha\mathrm{d}\lambda_{\alpha}\mathrm{d}\phi_{\alpha}\ .
\end{equation}
The joint distribution of eigenvalues is then obtained by integrating over the angle variables, using
\begin{equation}
\int_{-\infty}^{\infty}\exp \left [-B|V_{\alpha}|\cosh\mu\phi_{\alpha}\right ]\mathrm{d}\phi_{\alpha}=
\frac{1}{\mu}K_0\left(B|V_{\alpha}|\right)
\end{equation}
where $K_0$ is the modified Bessel function of the second kind. This yields the joint distribution of eigenvalues for model CM$_h$ as 
\begin{equation}
P(\lambda_1,\ldots, \lambda_N)\sim 
\exp\left(-A\sum_{\alpha}\lambda_{\alpha}^2\right)\prod_{\alpha}  K_0\left(B\prod_{\beta\neq \alpha}\left|1+\dfrac{\mathrm{i}g\mu}{\lambda_{\alpha}-\lambda_{\beta}}\right|\right)\ .
\label{distributionIInr}
\end{equation}
This expression is exact but difficult to handle. In order to find a Wigner-type surmise for the nearest-neighbour distributions we consider the limiting behaviour $P(\bfl)$ when two nearby eigenvalues $\lambda_1$ and $\lambda_2$ get close to each other. Setting $s=\lambda_1-\lambda_2$ we see that the factor $\exp(-b/s^2)$ in the case of model CM$_r$  is replaced by a factor
\begin{equation}
K_0\left(B\sqrt{1+\frac{g^2\mu^2}{s^2}}\right)^2\underset{s\to 0}{\sim}\ s\exp\left(- \frac{2B g\mu}{s}\right)\ .
\end{equation}
We therefore expect the nearest-neighbour spacing distribution to behave as 
\begin{equation}
P(n,s)=a s^d \exp(-b/s-c s)\ .
\label{surmiseIInr}
\end{equation}
In Fig.~\ref{psIInr} we show the results of numerical computations of the nearest-neighbour spacing distributions for matrices of the form \eqref{IInr} with the choice of parameters and variables detailed in section \ref{numericalimplementation}. The surmise \eqref{surmiseIInr} perfectly reproduces numerical results.

\begin{figure}[hbt]
\begin{center}
\includegraphics[width=.65\linewidth]{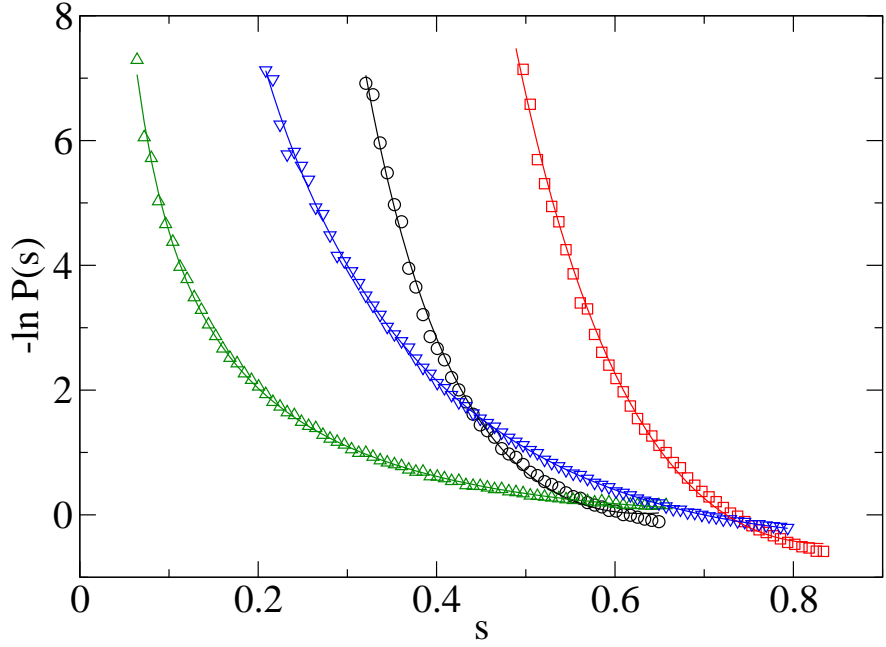}
\end{center}
\caption{(Color online). Nearest-neighbour spacing distributions $P(s)$ for the random matrices CM$_r$ ($g=0.5$, black circles) and $g=1$, red squares) and CM$_h$ ($g=.5$, green triangles up, and $g=1$, blue triangles down), with $\mu=4\pi/N$. Symbols are numerical results, solid lines indicate the fit $b/s^2+c s-\ln a$ (model CM$_r$, Eq.~\eqref{surmise}) and $b/s+c s-\ln a -d\ln s$ (model CM$_h$, Eq.~\eqref{surmiseIInr}). Logarithm is natural.  \label{psInr-IInr}}
\end{figure}

In order to assess better the validity of the exponentially strong level repulsion for models CM$_r$ and CM$_h$, we compare in Fig.~\ref{psInr-IInr} the beginning of the distributions $P(s)$ for these models. Clearly the $1/s^2$ repulsion for CM$_r$ and the $1/s$ repulsion for CM$_h$ fit numerical curves very well. However, the precision of our numerical results does not permit to confirm or reject the presence of the logarithmic term $d\ln s$ in $P(s)$ for CM$_h$ model.

%===================================================================================

\section{Trigonometric Calogero-Moser model}\label{model_CM_T}

For the trigonometric Calogero-Moser model CM$_t$ the Lax matrix is \cite{perelomov}
\begin{equation}
L_{kr}=p_r\delta_{kr}+\mathrm{i}g(1-\delta_{kr})\dfrac{\mu}{2\sin(\mu(q_k-q_r)/2)}\ .
\label{IIInr}
\end{equation}
The only difference with the previous model, (\ref{IInr}), is the $\sin$ function which replaces the $\sinh$. The matrix (\ref{IIInr}) can be obtained from (\ref{IInr}) by the substitution $\mu\to\mathrm{i}\mu$. However the fact that positions of the particles are now defined on a circle (because of the sin function) makes the resulting spectral statistics entirely different from the previous models CM$_r$ and CM$_h$.

To construct action-angle variable we introduce, as in the previous section, two matrices
\begin{equation}
Q_{\alpha \beta}=\sum_k u_k^*(\alpha)\mathrm{e}^{\mathrm{i}\mu q_k}u_k(\beta)\ ,\qquad 
R_{\alpha \beta}=\sum_k u_k^*(\alpha)\mathrm{e}^{-\mathrm{i}\mu q_k}u_k(\beta)\ ,
\label{Q_IIInr}
\end{equation}
and two vectors
\begin{equation}
e_{\alpha}=\sum_k u_k(\alpha)\mathrm{e}^{\mathrm{i}\mu q_k/2}\ ,\qquad 
f_{\alpha}=\sum_k u_k(\alpha)\mathrm{e}^{-\mathrm{i}\mu q_k/2}\ .
\label{e_alphaIIInr}
\end{equation}
Following the same steps as above with $\mu$ replaced by $\mathrm{i}\mu$, one gets
\begin{equation}
Q_{\alpha \beta}=f_{\alpha}^*\dfrac{g\mu}{\lambda_{\alpha}-\lambda_{\beta}+g\mu }e_{\beta}\ ,\qquad
R_{\alpha \beta}=e_{\alpha}^*\dfrac{g\mu}{\lambda_{\beta}-\lambda_{\alpha}+g\mu }f_{\beta}\ .
\label{Q_IIInr_new}
\end{equation}
Again, using the fact that $Q$ is the inverse of $R$ we obtain that 
\begin{equation}
\label{equationefIIInr}
\sum_{\gamma}\frac{|e_{\gamma}|^2}{\lambda_{\gamma}-\lambda_{\alpha}-g\mu}=c_1\ , 
\qquad
\sum_{\gamma}\frac{|f_{\gamma}|^2}{\lambda_{\gamma}-\lambda_{\alpha}+g\mu}=c_2
\end{equation}
with certain constants $c_1$ and $c_2$ independent on $\alpha$. Repeating the same arguments as in the previous section and using results of Appendix~\ref{app_A} one concludes that $c_2=-c_1=1/(\mu g)$ and 
\begin{equation}
|e_{\alpha}|^2=V_{\alpha}\ ,\qquad |f_{\alpha}|^2=W_{\alpha}\ ,
\label{efW}
\end{equation}
with
\begin{equation}
V_{\alpha}=\prod_{\beta\neq \alpha}\left(1-\dfrac{g\mu}{\lambda_{\alpha}-\lambda_{\beta}}\right)\ ,\qquad
W_{\alpha}=\prod_{\beta\neq \alpha}\left(1+\dfrac{g\mu}{\lambda_{\alpha}-\lambda_{\beta}}\right)\ .
\end{equation}
The new variables $\phi_{\alpha}$ are  defined  as above from the diagonal elements of matrix $Q$ as follows 
\begin{equation}
Q_{\alpha \alpha}=V_{\alpha}^{1/2}W_{\alpha}^{1/2}\mathrm{e}^{\mathrm{i}\mu \phi_{\alpha}}\ .
\label{phase_Qaa}
\end{equation} 
In the definition \eqref{e_alphaIIInr} of $e_{\alpha}$ one has a freedom to choose the overall phase of the eigenvector $u_k(\alpha)$. Since from \eqref{Q_IIInr_new} one must have $Q_{\alpha \alpha}=f_{\alpha}^*e_{\alpha}$, one can choose phases, for example, as follows 
\begin{equation}
e_{\alpha}=V_{\alpha}^{1/2}\mathrm{e}^{\mathrm{i}\mu \phi_{\alpha}/2}\ ,\qquad
f_{\alpha}=W_{\alpha}^{1/2}\mathrm{e}^{-\mathrm{i}\mu \phi_{\alpha}/2}\ .
\label{resultat_eIIInr}
\end{equation}
Then the matrix $Q$ can be expressed in terms of new variables $\lambda_{\alpha}$ and $\phi_{\alpha}$ as
\begin{equation}
Q_{\alpha \beta}=\mathrm{e}^{\mathrm{i}\mu \phi_{\alpha}/2}W_{\alpha}^{1/2}
\dfrac{g\mu}{\lambda_{\alpha}-\lambda_{\beta}+g\mu}
V_{\beta}^{1/2}\mathrm{e}^{\mathrm{i}\mu \phi_{\beta}/2}\ .
\label{QIIInr}
\end{equation}
The inverse matrix $R$ plays a symmetric role, as it is obtained from $Q$ by exchanging $\mu$ to $-\mu$. Again, there is a canonical transformation from position and momentum variables $(q_k,p_k)$ to action and angle variables $(\lambda_{\alpha},\phi_{\alpha})$ \cite{I}.

An important consequence of \eqref{efW} is that for all $\alpha$ we should have 
\begin{equation}
V_{\alpha}> 0,\qquad W_{\alpha}> 0\ .
\label{conditions_III}
\end{equation}
These inequalities impose non-trivial restrictions on eigenvalues $\lambda_{\alpha}$, as we will see now. We label eigenvalues so that $\lambda_1< \lambda_2<\ldots < \lambda_N$, and we consider the function
\begin{equation}
h(x)=\sum_{\gamma}\frac{V_{\alpha}}{\lambda_{\gamma}-x}+\frac{1}{\mu g}\ .
\end{equation} 
It has $N$ poles at $x=\lambda_{\gamma}$, and according to Eqs.~\eqref{equationefIIInr} and \eqref{resultat_eIIInr} it has $N$ zeros at $x=\lambda_{\alpha}+\mu g$. If all numerators $V_{\alpha}$ are positive then the derivative of $h$ is positive, and it is easy to check from the graph of the function that between two consecutive poles there is one and only one zero. Suppose $\mu g>0$. Then for $x\to -\infty$ the function $h(x)$ has a strictly positive limit. The lowest zero $\lambda_1+\mu g$ must thus lie in the interval $]\lambda_1,\lambda_2[$. More generally one must have $\lambda_{\alpha}+\mu g\in ]\lambda_{\alpha},\lambda_{\alpha+1}[$ for $1\leq\alpha\leq N-1$, while the largest zero $\lambda_N+\mu g$ lies in the interval $]\lambda_N,\infty[$. Thus eigenvalues fulfill the inequalities
\begin{equation}
\label{difflambda}
\lambda_{\alpha+1}-\lambda_{\alpha}> g\mu\ .
\end{equation}
Conversely, if these inequalities are fulfilled then trivially all $V_{\alpha}$ are positive. Therefore, \eqref{difflambda} are the necessary and sufficient conditions for the positivity of all $V_{\alpha}$. In particular, eigenvalues of the Lax matrix \eqref{IIInr} obey inequalities \eqref{difflambda} for any choice of the $\bfq$. These results adapt straightforwardly to the case where $\mu g$ is negative. 

Since in \eqref{IIInr} the $q_k$ only appear as an argument in the $\sin$ function, there is no need to choose a confining potential for the particle distribution as in models CM$_r$ and CM$_h$. We consider the probability distribution of $\bfp$ and $\bfq$ in the form 
\begin{equation}
P(\bfp\, ,\bfq\, ) \sim \exp \left [-A\Big (\sum_j p_j^2+g^2\sum_{j\neq k}\frac{\mu^2}{4\sin^2 \big [\mu(q_k-q_r)/2\big ] } \Big ) \right ]
\end{equation} 
with the restrictions that all $q_j$ are between $0$ and $2\pi/\mu$. Since the change of variables from $\bfp$ and $\bfq$ to $\lambda_{\alpha}$ and $\phi_{\alpha}$ is canonical and the restrictions \eqref{difflambda} do not depend on phase variables the joint distribution of eigenvalues is 
\begin{equation}
P(\lambda_1,\ldots, \lambda_N) \sim \exp\left(-A\sum_{\alpha}\lambda_{\alpha}^2\right) \chi(\lambda_1,\ldots,\lambda_N)\ ,
\end{equation}
where the function $\chi(\bfl)$ is equal to 1 if \eqref{difflambda} is fulfilled for all $\alpha$, and 0 otherwise.  

\begin{figure}[hbt]
\begin{center}
\includegraphics[width=.65\linewidth]{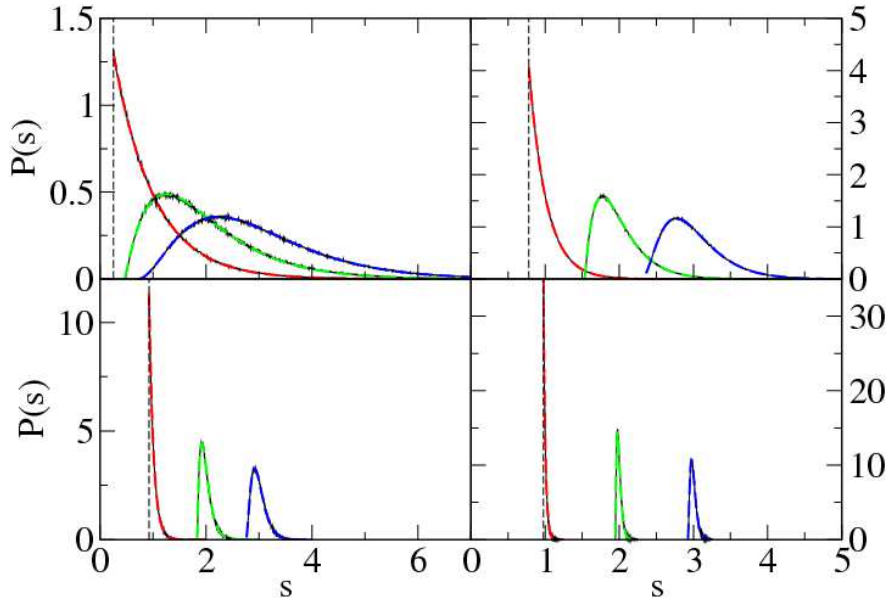}
\end{center}
\caption{(Color online). Same as Fig.~\ref{psIInr} for model CM$_t$ with $P(n,s)$ the shifted Poisson distribution~\eqref{pnsIIInr}.\label{psIIInr}}
\end{figure}

It turns out that model CM$_t$ is very similar to a fourth model, the Ruijsenaars-Schneider model, that we will consider in the next section. Therefore we postpone analytical calculations of the nearest-neighbour spacing distributions to section~\ref{transfer_op}. The nearest-neighbour spacing distributions $P(n,s)$ are shifted Poisson distributions of the form 
\begin{equation}
\label{pnsIIInr}
P(n,s)=\left \{ \begin{array}{ll} 0,& 0<s<nb \\ \dfrac{(s-n b)^{n-1}}{(n-1)!(1-b)^n} \mathrm{e}^{- (s-nb)/(1-b)}, & s>nb \end{array} \right . 
\end{equation}
with some numerical constant $b$. In Fig.~\ref{psIIInr} we show the results of numerical computations for matrices of the form \eqref{IIInr} with the choice of parameters and variables detailed in section \ref{numericalimplementation}.  

%=============================================================================
\section{Ruijsenaars-Schneider model}\label{model_RS}

The Calogero-Moser models considered in the previous sections are such that there exists a matrix $Q$ which is, in a certain sense,  dual to the Lax matrix $L$. Namely, the canonical transformation from variables $(p_k,q_k)$ to action-angle variables $(\lambda_{\alpha}, \phi_{\alpha})$ is such that the action variables $\lambda_{\alpha}$ are eigenvalues of $L$ and angle variables $\phi_{\alpha}$ are related to $Q_{\alpha \alpha}$ in a simple way. In fact, the matrix $Q$ is also a Lax matrix, corresponding to a possibly different Hamiltonian, and $L$ plays the role of a matrix dual to $Q$ for the inverse of the canonical transformation  \cite{I}. The rational Calogero-Moser system is self-dual since matrices $L$ and $Q$, given by \eqref{Inr} and \eqref{QInr}, are equal up to labelling of the variables. 

The model we consider in this section is related to the above models in that its Lax matrix $L$ and the dual matrix $Q$ are a kind of generalisation of those of model CM$_t$ \eqref{QIIInr}. The treatment of this model closely follows the previous section. 

The Lax matrix for Ruijsenaars-Schneider model  is the $N\times N$ unitary matrix given by \cite{III}
\begin{equation}
L_{kr}=\mathrm{e}^{\mathrm{i}\sigma p_k/2}\tilde{W}_k^{1/2}
\dfrac{\sin \big [\mu g\sigma/2\big ] }{\sin \big [\mu (q_k-q_r+g\sigma)/2\big ]}
\tilde{V}_r^{1/2}\mathrm{e}^{\mathrm{i}\sigma p_r/2}\ .
\label{IV}
\end{equation}
with
\begin{equation}
\label{VWtildeIV}
\tilde{V}_k=\prod_{j\neq k}\frac{\sin \big [\mu (q_k-q_j-g\sigma)/2\big ]}{\sin \big [\mu (q_k-q_j)/2\big ]}\ ,\qquad
\tilde{W}_k=\prod_{j\neq k}\frac{\sin \big [\mu (q_k-q_j+g\sigma)/2\big ]}{\sin \big [\mu (q_k-q_j)/2\big ]}\ .
\end{equation}
This Lax matrix is related with the Hamiltonian \eqref{hamiltonian_rs} by
\begin{equation}
H(\bfp,\bfq)=\frac{1}{2}\mathrm{Tr}(L+L^{\dag})\ .
\end{equation}
It is convenient to introduce the vectors
\begin{equation}
\tilde{e}_k=\tilde{V}_k^{1/2}\mathrm{e}^{\mathrm{i}\sigma p_{k}/2}\mathrm{e}^{\mathrm{i}\mu q_k/2}\ ,\qquad
\tilde{f}_k=\tilde{W}_k^{1/2}\mathrm{e}^{\mathrm{i}\sigma p_{k}/2}\mathrm{e}^{-\mathrm{i}\mu q_k/2}\ ,
\label{eIVtilde}
\end{equation} 
so that matrix $L$ can be rewritten as
\begin{equation}
L_{kr}=\tilde{f}_k^*\mathrm{e}^{-\mathrm{i}\mu q_k/2}
\dfrac{\sin \big [\mu g\sigma/2\big ]}{\sin \big [\mu (q_k-q_r+g\sigma)/2\big ]}
\mathrm{e}^{-\mathrm{i}\mu q_r/2}\tilde{e}_r\ .
\label{LefIV}
\end{equation}
As in the previous section, the condition that the matrix \eqref{IV} is unitary imposes certain restrictions on the coordinates $\bfq$, which we will discuss later. Assuming that the Lax matrix is unitary, we choose to denote its eigenvalues by $\mathrm{e}^{\mathrm{i}\sigma \lambda_{\alpha}}$. The dual matrices for the Ruijsenaars-Schneider model  are defined by \cite{III} 
\begin{equation}
Q_{\alpha \beta}=\sum_k u_k^*(\alpha)\mathrm{e}^{\mathrm{i}\mu q_k}u_k(\beta)\ ,\qquad
R_{\alpha \beta}=Q_{\beta \alpha }^*=\sum_k u_k^*(\alpha)\mathrm{e}^{-\mathrm{i}\mu q_k}u_k(\beta)\ ,
\label{Q_IV}
\end{equation}
and the vectors $e_{\alpha}$ and $f_{\alpha}$ by
\begin{equation}
e_{\alpha}=\sum_k u_k(\alpha)\tilde{e}_k\ ,\qquad f_{\alpha}=\sum_k u_k(\alpha)\tilde{f}_k\ .
\label{e_alphaIV}
\end{equation}
From Eq.~\eqref{LefIV} one has 
\begin{equation}
\mathrm{e}^{\mathrm{i}\mu(q_k+q_r)/2} L_{kr}\sin  \big [\mu (q_k-q_r+g\sigma)/2\big ]=\tilde{f}_k^*\sin \big [\mu g\sigma/2\big ] \tilde{e}_r\ . 
\end{equation}
Multiplying this expression by $u_k^*(\alpha)u_r(\beta)$ and summing both sides over $k$ and $r$ leads to
\begin{equation}
\frac{1}{2\mathrm{i} }Q_{\alpha \beta}\left ( \mathrm{e}^{\mathrm{i} \sigma \lambda_{\beta}+\mathrm{i}\sigma\mu g/2} -
\mathrm{e}^{\mathrm{i} \sigma \lambda_{\alpha}-\mathrm{i}\sigma\mu g/2}\right )=f_{\alpha}^*\sin \big [\mu g\sigma/2\big ] e_{\beta}\ ,
\end{equation}
which yields the analogue of \eqref{Q_IIInr_new},
\begin{equation}
Q_{\alpha \beta}=f_{\alpha}^*\mathrm{e}^{-\mathrm{i}\sigma\lambda_{\alpha}/2}
\dfrac{\sin  \big[\sigma g\mu/2\big ]}{\sin \big[\sigma (\lambda_{\beta}-\lambda_{\alpha}+g\mu)/2 \big ]}
\mathrm{e}^{-\mathrm{i}\sigma\lambda_{\beta}/2}e_{\beta}\ .
\label{Q_IV_new}
\end{equation}
Let us rewrite 
\begin{equation}
Q_{\alpha\beta}=f_{\alpha}^*
\dfrac{\mathrm{e}^{-\mathrm{i}\tau}-1}
{\mathrm{e}^{\mathrm{i}(\sigma \lambda_\alpha-\tau)}-
\mathrm{e}^{\mathrm{i}\sigma \lambda_\beta}}e_{\beta}\ ,
\label{QIVsimple}
\end{equation}
where we have set $\tau=\sigma\mu g$. From its definition~\eqref{Q_IV} it is clear that $Q$ has to be an unitary matrix, i.e.~$\sum_{\gamma} Q_{\alpha \gamma}Q_{\beta \gamma}^*=\delta_{\alpha \beta}$. Selecting terms with $\alpha=\beta$ and $\alpha\neq \beta$ yields the two equations
\begin{eqnarray}
|\rho|^2|f_{\alpha}|^2\sum_{\gamma}\frac{|e_{\gamma}|^2}{|\mathrm{e}^{\mathrm{i}(\sigma \lambda_\alpha-\tau)}-\mathrm{e}^{\mathrm{i}\sigma \lambda_\gamma}|^2}&=&1\ ,
\label{conditionsQR1-IIInr}\\
\sum_{\gamma}\frac{|e_{\gamma}|^2}{(\mathrm{e}^{\mathrm{i}(\sigma \lambda_\alpha-\tau)}-\mathrm{e}^{\mathrm{i}\sigma \lambda_\gamma})(\mathrm{e}^{-\mathrm{i}(\sigma \lambda_\beta-\tau)}-\mathrm{e}^{-\mathrm{i}\sigma \lambda_\gamma})}&=&0\ ,
\label{conditionsQR2-IIInr}
\end{eqnarray}
where we have set $\rho=\mathrm{e}^{-\mathrm{i}\tau}-1$. Using the identity
\begin{eqnarray}
& &\frac{1}{(\mathrm{e}^{\mathrm{i}(\sigma \lambda_\alpha-\tau)}-\mathrm{e}^{\mathrm{i}\sigma \lambda_\gamma})(\mathrm{e}^{-\mathrm{i}(\sigma \lambda_\beta-\tau)} -\mathrm{e}^{-\mathrm{i}\sigma \lambda_\gamma})}=
\nonumber\\
& &\frac{\mathrm{e}^{\mathrm{i}(\sigma \lambda_\beta-\tau)}}
{\mathrm{e}^{\mathrm{i}(\sigma \lambda_\alpha-\tau)}-\mathrm{e}^{\mathrm{i}(\sigma \lambda_\beta-\tau)}}
\left(\frac{\mathrm{e}^{\mathrm{i}\sigma \lambda_\gamma }}{\mathrm{e}^{\mathrm{i}(\sigma \lambda_\alpha-\tau)}-\mathrm{e}^{\mathrm{i}\sigma \lambda_\gamma}}+\frac{\mathrm{e}^{\mathrm{i}\sigma \lambda_\gamma}}{\mathrm{e}^{\mathrm{i}\sigma \lambda_\gamma}-\mathrm{e}^{\mathrm{i}(\sigma \lambda_\beta-\tau)}}\right)\ ,
\end{eqnarray}
it follows from \eqref{conditionsQR2-IIInr}  that there exists a constant $c$ such that 
\begin{equation}
\label{eqvpIV}
\sum_{\gamma}\frac{|e_{\gamma}|^2\mathrm{e}^{\mathrm{i}\sigma \lambda_\gamma}}{\mathrm{e}^{\mathrm{i}\sigma \lambda_\gamma}-\mathrm{e}^{\mathrm{i}(\sigma \lambda_\alpha-\tau)}}=c\ .
\end{equation}
Equations \eqref{eqn}--\eqref{bm} allow to obtain
\begin{equation}
|e_{\alpha}|^2=-c\rho V_{\alpha}\mathrm{e}^{-\mathrm{i}\tau(N-1)/2}\ ,
\end{equation}
while from Eq.~\eqref{conditionsQR1-IIInr} one obtains, using Eq.~\eqref{sum_bm2}
\begin{equation}
|f_{\alpha}|^2=-\frac{1}{c\rho} W_{\alpha}\mathrm{e}^{\mathrm{i}\tau(N-1)/2}\ ,
\end{equation} 
with new vectors $V_{\alpha}$ and $W_{\alpha}$ defined by
\begin{equation}
V_{\alpha}=\prod_{\beta\neq \alpha}\frac{\sin \big [\sigma (\lambda_{\alpha}-\lambda_{\beta}+g\mu)/2\big ]}{\sin \big [ \sigma (\lambda_{\alpha}-\lambda_{\beta})/2\big ]}\ ,\qquad 
W_{\alpha}=\prod_{\beta\neq \alpha}\frac{\sin\big [\sigma (\lambda_{\alpha}-\lambda_{\beta}-g\mu)/2\big ]}{\sin\big [ \sigma (\lambda_{\alpha}-\lambda_{\beta})/2\big ]}\ .
\label{VWIV}
\end{equation}
Using \eqref{sum_norm} we get
\begin{equation}
\sum_{\alpha}|e_{\alpha}|^2=c(1-\mathrm{e}^{-\mathrm{i}\tau N})\ .
\end{equation}
There is some overall freedom in the definition of vectors $\tilde{e}_k$ and  $\tilde{f}_k$ in Eq.~\eqref{eIVtilde}, as one could multiply $\tilde{e}_{\alpha}$ by some constant factor and divide $\tilde{f}_{\alpha}$ by the same factor. This in turn entails the same freedom for vectors $e_{\alpha}$ and  $f_{\alpha}$ in \eqref{e_alphaIV}. If one chooses for instance $\sum_j |\tilde{e}_j|^2=t$ then from unitarity of the transformation in \eqref{e_alphaIV} one has $\sum_{\alpha}|e_{\alpha}|^2=t$, which fixes the value 
\begin{equation}
c=\frac{t}{1-\mathrm{e}^{-\mathrm{i}\tau N}}
\label{c_RS}
\end{equation}
and thus 
\begin{equation}
|e_{\alpha}|^2=\frac{t \sin(\tau/2)}{\sin(N\tau/2)} V_{\alpha}\ ,\qquad  
|f_{\alpha}|^2=\frac{\sin (N\tau/2)}{t \sin(\tau/2)} W_{\alpha}\ .
\end{equation} 
By definition $t=\sum_j |\tilde{e}_j|^2$ is positive. A convenient choice is to take 
\begin{equation}
\label{choicet}
t=\left |\frac{\sin(N\tau/2)}{\sin(\tau/2)}\right |\ .
\end{equation}
Since $|e_{\alpha}|^2$ and $|f_{\alpha}|^2$ are non-negative one concludes that $V_{\alpha}$ and $ W_{\alpha}$ have the same sign as $\sin(N\tau/2)/\sin(\tau/2)$ for all $\alpha$. The choice \eqref{choicet} implies that
\begin{equation}
|e_{\alpha}|^2=|V_{\alpha}|\ ,\qquad |f_{\alpha}|^2=|W_{\alpha}|\ .
\label{efVWIV}
\end{equation}
The new action variables are the $\lambda_{\alpha}$, and angle variables $\phi_{\alpha}$ are defined by 
\begin{equation}
\label{defphiIV}
Q_{\alpha \alpha}=(V_{\alpha}W_{\alpha})^{1/2}\mathrm{e}^{\mathrm{i}\mu \phi_{\alpha}}\ .
\end{equation} 
Equation \eqref{Q_IV_new} implies that $Q_{\alpha\alpha}=f_{\alpha}^*e_{\alpha}\mathrm{e}^{-\mathrm{i}\sigma\lambda_{\alpha}}$. In view of \eqref{efVWIV} and \eqref{defphiIV} one can choose the phases of the eigenvectors $u_k(\alpha)$ such that $e_{\alpha}$ and $f_{\alpha}$ defined by \eqref{e_alphaIV} can be expressed as
\begin{equation}
e_{\alpha}=V_{\alpha}^{1/2}\mathrm{e}^{\mathrm{i}\mu \phi_{\alpha}/2}\mathrm{e}^{\mathrm{i}\sigma\lambda_{\alpha}/2}\ ,\qquad 
f_{\alpha}=W_{\alpha}^{1/2}\mathrm{e}^{-\mathrm{i}\mu \phi_{\alpha}/2}\mathrm{e}^{-\mathrm{i}\sigma\lambda_{\alpha}/2}\ .
\label{resultat_eIV}
\end{equation} 
In terms of the new variables, $Q$ thus reads
\begin{equation}
Q_{\alpha \beta}=\mathrm{e}^{\mathrm{i}\mu \phi_{\alpha}/2}W_{\alpha}^{1/2}
\dfrac{\sin \big [\sigma g\mu/2\big ]}{\sin \big [\sigma (\lambda_{\beta}-\lambda_{\alpha}+g\mu)/2\big ]}
V_{\beta}^{1/2}\mathrm{e}^{\mathrm{i}\mu \phi_{\beta}/2}\ .
\label{QIV}
\end{equation}
Comparing \eqref{IV} and \eqref{QIV} one sees that for  model RS  the dual matrix  matrix $Q$ is obtained from $L$ by changing $\sigma\leftrightarrow \mu$, $g\to -g$, $\bfp \to \bfph$, and $\bfq\to \bfl$. It means that this model  is self-dual (the matrices $L$ and $Q$ are the same up to a change of notation). One can show that the transformation from $(q_k,p_k)$ to action and angle variables $(\lambda_{\alpha},\phi_{\alpha})$ is canonical \cite{III}. As mentioned, a consequence of the unitarity of the matrix $Q$ is that $V_{\alpha}$ and $ W_{\alpha}$ have the same sign as $\sin(N\tau/2)/\sin(\tau/2)$ for all $\alpha$. This implies that certain inequalities have to be verified by the $\lambda_{\alpha}$, which we now derive in a way similar as in the previous section. 

Let us define the function 
\begin{equation}
h(x)=\sum_{\gamma}V_{\gamma}\cot\big [(x-\sigma \lambda_\gamma)/2\,]-t\cot (N\tau/2\,)\ ,
\end{equation}
which is periodic with period $2\pi$ and can be considered as a function on the unit circle. It has $N$ poles at $x=\sigma \lambda_{\gamma}$. Taking the imaginary part of Eq.~\eqref{eqvpIV}, with $c$ given by~\eqref{c_RS} and $e_{\alpha}$ given by \eqref{resultat_eIV}, one obtains that $h(x)$ has $N$ zeros at  $x=\sigma\lambda_\gamma -\tau$. When all $V_{\gamma}$ are positive, the same arguments as  in the previous section imply that between two consecutive poles of $h(x)$ there must be exactly one zero. It means that between two nearby eigenvalues $\sigma \lambda_\alpha$ there is one and only one number of the form $\sigma \lambda_\gamma-\tau$. A similar reasoning starting from matrix $R=Q^{\dag}$ leads to the conclusion that between two consecutive $\sigma \lambda_\gamma$ there must also be one and only one number of the form $\sigma \lambda_\alpha +\tau$. These two conditions (shift by $+\tau$ or $-\tau$) are equivalent, thus we can restrict ourselves to a shift by $+\tau$, i.e.~the condition that the sets $\{\sigma \lambda_{\gamma}, 1\leq \gamma\leq N\}$ and $\{\sigma \lambda_{\gamma}+\tau, 1\leq \gamma\leq N\}$ intertwine on the unit circle. It is a necessary and sufficient condition for  the matrix \eqref{Q_IV_new} to be an unitary matrix. A similar conclusion is readily obtained when all $V_{\gamma}$ are negative. 

The conditions implied by this type of intertwining have been discussed in \cite{Schmit,Remy}. There is an fundamental difference between these eigenvalue conditions in the RS model and in the CM$_t$ model discussed in the previous section. In the CM$_t$ case the Lax matrix is Hermitian, and the $\lambda_\alpha$ can take values on the whole real axis. In the RS case, as the Lax matrix is unitary, the $\sigma\lambda_\alpha$ lie on the unit circle. Therefore poles and zeros cannot be ordered in a simple way as in the previous case, and the analysis of the previous section does not apply.

\begin{figure}[ht]
\begin{center}
\includegraphics[width=.3\linewidth]{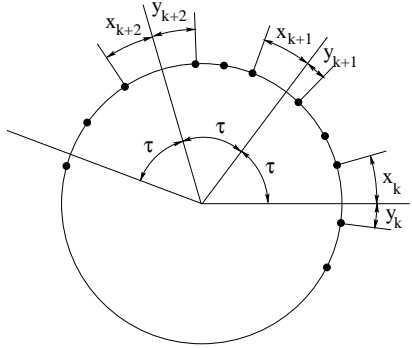}
\end{center}
\caption{Division of the circle into sectors of angle $\tau$. Only sectors numbers $k$, $k+1$, and $k+2$ are indicated. Black circles indicate the positions of the $\sigma\lambda_\alpha$, where $\exp(\mathrm{i}\sigma\lambda_\alpha)$ are the eigenvalues of the RS Lax matrix.}
\label{sectors}
\end{figure}

 For completeness we shortly repeat the arguments of the papers~\cite{Schmit,Remy}. Let us put all eigenvalues  $\sigma \lambda_\alpha$ of the Lax matrix \eqref{IV} on the unit circle and divide the circle into  sectors with angle $\tau$.  Denote the (positive) angular distance from the boundaries of the $k$th sector in counter clockwise direction by $x_k$ and in clockwise direction by $y_k$, as in Fig.~\ref{sectors}.
After a shift by $\tau$, the intertwining relations imply that only one of the two points corresponding to $x_k$ and $y_k$ will fall in-between points corresponding to $x_{k+1}$ and $y_{k+1}$. The first case corresponds to $x_{k}>x_{k+1}$ and $y_{k+1}>y_k$. In the second case the inequalities are reversed and $x_{k}<x_{k+1}$ and $y_{k+1}<y_k$. In both cases the inequality
\begin{equation}
(y_{k+1}-y_k)(x_{k+1}-x_{k})<0
\label{y_x}
\end{equation}
is fulfilled.  

Let us consider consecutive sectors of angle $\tau$ as in Fig.~\ref{sectors}. Denote the number of eigenvalues in each sector by $n_k$. After a shift by $\tau$, eigenphases from the $k$th sector will move into the $(k+1)$th sector. The $n_k$ shifted points divide this sector into $n_k+1$ intervals. As was proved above, eigenphases in the $(k+1)$th sector have to intertwine with these shifted eigenphases. Therefore all  $n_k+1$ intervals except the first and the last will be occupied. The first will be occupied if $x_{k+1}<x_k$ and the last interval will be occupied provided $y_{k+1}>y_{k+2}$. These statements can be rewritten in the form of the recurrence  relation 
\begin{equation}
n_{k+1}=n_k-1+ \Theta(x_{k}-x_{k+1})+\Theta(y_{k+1}-y_{k+2})\ ,
\end{equation}
where $\Theta(t)$ is the Heaviside step function, $\Theta(t)=1$ when $t>0$ and  $\Theta(t)=0$ for $t<0$. From  \eqref{y_x} it follows that this relation can be rewritten in the form
\begin{equation}
n_{k+1}=n_k-1+ \Theta(x_{k}-x_{k+1})+\Theta(x_{k+2}-x_{k+1})\ .
\label{nknk}
\end{equation} 
We now specialise to the case where $\tau$ depends on the size $N$ of the Lax matrix. We set
\begin{equation}
\tau=\frac{2\pi}{N}a
\label{tau_RS}
\end{equation}
with fixed $a$. The case of integer $a$ is trivial and we thus assume that $a$ is not an integer. The total number of sectors of angle $\tau$ in the unit circle is
\begin{equation}
K=\left[ \frac{N}{a}\right]
\end{equation}
where $[t]$ denotes the integer part of $t$. Suppose the beginning of the first sector lies at position $\sigma\lambda_1$. We choose to consider that this eigenvalue does not belong to the first sector, i.e.~$y_1=0$. Applying Eq.~\eqref{y_x} for $k=0$ implies that $x_1>x_0$. Thus necessarily $n_2=n_1+1$ and for $k\geq 3$ one easily gets from Eq.~\eqref{nknk} that 
\begin{equation}
\label{secteurs}
n_k=n_1+\Theta(x_{k+1}-x_{k})\ .
\end{equation}  The total number of eigenvalues lying into all  $K$ sectors  obeys the inequalities 
\begin{equation}
N-n_1-1\leq \sum_{k=1}^K n_k\leq N-1\ .
\label{first_inequality}
\end{equation}
The right-hand side inequality comes from the fact that when $a$ is not an integer the union of all $K$ intervals does not overlap the whole circle: in particular, it does not contain the first eigenvalue $\sigma\lambda_1$ from which we start our sectors. The left-hand side inequality is a consequence of the fact that if eigenvalues were shifted  by $-\tau$, the first sector in the opposite direction would have exactly the same number of eigenvalues as the second sector, i.e. $n_1+1$ eigenvalues: as all $K$ sectors does not cover the whole circle and the uncovering region is smaller than the sector of angle $\tau$, it follows that the number of eigenvalues in all $K$ sectors is larger than $N-(n_1+1)$. 

From \eqref{secteurs} one easily obtains a second inequality
\begin{equation}
Kn_1+1\leq \sum_{k=1}^K n_k\leq K(n_1+1)\ .
\label{second_inequality}
\end{equation}
From these two inequalities it follows that
\begin{equation}
\frac{N}{K+1}-1 \leq n_1\leq \frac{N-2}{K}\ .
\label{n_1}
\end{equation}
By definition of $K$ we have
\begin{equation}
\frac{N}{a}-1<K<\frac{N}{a}\ . 
\end{equation}
Substituting in the right-hand side of \eqref{n_1} the minimum of $K$ and in the left-hand side the maximum value of $K$ one gets  
\begin{equation}
a-1-\frac{a^2}{N+a}<n_1<a+\frac{a(a-2)}{N-a}\ ,
\end{equation}
which entails that for $N$ large enough $a-1<n_1<a$. Since $n_1$ has to be an integer, $n_1=[a]$. This means that for sufficiently large $N$, within an interval of length $\tau$ from any eigenvalue there are always exactly $[a]$ eigenvalues.

The inverse statement is also true. If within an interval of length $\tau$ from any eigenvalue there exist exactly $[a]$ other eigenvalues then all $V_{\alpha}$ and $W_{\alpha}$ have the sign of $\sin (\tau/2)/\sin(N\tau/2)$. To see this let us consider e.g.~$V_{\alpha}$ given by Eq.~\eqref{VWIV}. It is a product of terms $\sin x / \sin (x-\tau)$ where $x=\sigma (\lambda_{\alpha}-\lambda_{\beta})/2$. As the distance between two eigenvalues may be restricted, $0<\sigma \lambda_{\alpha}-\sigma\lambda_{\beta}<2\pi $,  $x$ obeys inequality $0<x<\pi$. Therefore $\sin(x)>0$, and $\sin (x-\tau)$ is negative when $0<x<\tau$ and positive when $\tau<x<\pi$. If within an interval of length $\tau$ from $\sigma\lambda_{\alpha}$ there are exactly $[a]$ eigenvalues, then in the product formula for $V_{\alpha}$ there are exactly $[a]$ negative terms, so that its total sign is $(-1)^{[a]}$. For $N$ large enough the sign of $\sin (\tau/2)/\sin(N\tau/2)$ with $\tau=2\pi a /N$ is the sign of $\sin\pi a$, which is precisely $(-1)^{[a]}$.

The above arguments prove that for sufficiently large $N$ (whose value depends only on $a$) the necessary and sufficient condition for the unitarity of matrix $Q$ is that at distance $|\tau|$ from any eigenvalue there exist $[a]$ other eigenvalues. 

As matrices $L$ and $Q$ are dual, the unitarity condition for $L$ can be readily deduced: it is that at distance $|\tau|$ from one coordinate $\mu q_k$ there are exactly $[a]$ other coordinates. Note that in \cite{III} only the case $0<a<1$ had been considered.  

These restrictions determine the allowed region in coordinate space. We choose as the 'natural' measure of momenta and coordinates the uniform distribution for momenta (between $0$ and $2\pi/\sigma$) and coordinates uniformly distributed in the allowed region as explained above. After the change of variables from coordinates and momentum to action-angle variables it follows that the resulting distribution of eigenvalues will be also uniform but in the allowed region of eigenvalues 
with the only restriction that any interval $]\sigma \lambda_\alpha,\sigma \lambda_\alpha+2\pi a/N[$ contains exactly $[a]$ eigenvalues. In next section we will show that it is possible to calculate asymptotic expressions for the joint distribution of eigenvalues by using a transfer operator technique.

For numerical investigations, we consider an ensemble of unitary matrices of the form \eqref{LefIV}, with $p_k$ chosen as independent random variables uniformly distributed between $0$ and $2\pi$ and the picket fence distribution of coordinates $q_j=j$, $1\leq j\leq N$ (see section \ref{numericalimplementation}). Choosing constants such that $\mu=2\pi/N$, $\sigma=1$, and $g=a$, so that $\tau=2\pi a/N$, direct calculations yield
\begin{equation}
\tilde{V}_k=\tilde{W}_k=\left|\frac{\sin N\tau/2}{N\sin\tau/2}\right|\ . 
\end{equation}
With a slightly different choice of phases in \eqref{IV}, matrix $L$ simplifies to
\begin{equation}
L_{kr}=\frac{\mathrm{e}^{\mathrm{i}\Phi_k}}{N}\frac{1-\mathrm{e}^{2\mathrm{i}\pi a}}{1-\mathrm{e}^{2\mathrm{i}\pi((k-r+a)/N)}}\ ,
\label{pseudo_map}
\end{equation}
where we denote $p_k=\Phi_k$. This is a particular specialisation of the Lax matrix for the Ruijsenaars-Schneider model. In the form \eqref{pseudo_map} but with $a=b N$ with fixed $b$ it first appeared in \cite{Schmit} as a result of the quantisation of a classical parabolic map on the torus proposed in \cite{giraud}.  When $b$ is a rational number the map considered in \cite{giraud} corresponds to a pseudo-integrable map of exchange of two intervals. For this particular case where $a$ depends on $N$, the spectral statistics of the unitary matrix \eqref{pseudo_map} has been obtained analytically in \cite{Schmit} and \cite{Remy} without knowledge of the relation with the Lax matrix of the  Ruijsenaars-Schneider model. In the present case, where $a$ is a fixed parameter independent on $N$, results are quite different. Analytical calculations of spectral correlation functions for this model are performed in the next sections.

\begin{figure}[ht]
\begin{center}
\includegraphics[width=.65\linewidth]{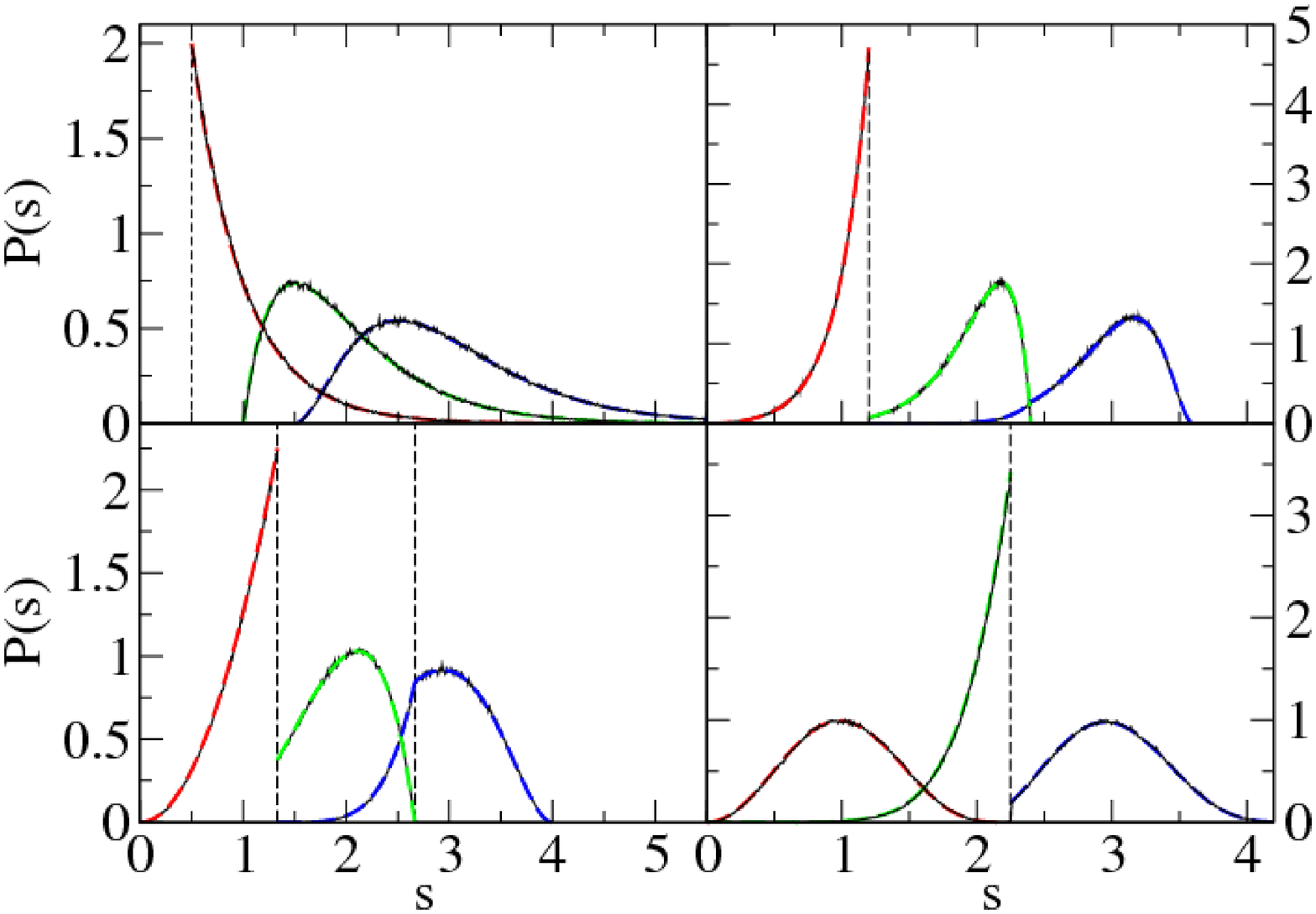}
\end{center}
\caption{Nearest-neighbour distributions $P(n,s)$ for the random matrices of model RS with $\mu=2\pi/N$ and $g=1/2$ (top left), $6/5$ (top right), $4/3$ (bottom left), and $9/4$ (bottom right), averaged over $1000$ realisations of matrices of size $N=701$. Solid lines are numerical results, dashed lines indicate the analytical expressions \eqref{pns01} for $g=1/2$, (\ref{nearest})--(\ref{p3s}) for $g=1.2$ and $g=4/3$, and \eqref{p1s23}--\eqref{p3s23} for $g=9/4$. In each panel (from left to right) $P(s)=P(1,s)$ (red), $P(2,s)$ (green) and $P(3,s)$ (blue). \label{psIV}}
\end{figure}

 To illustrate the accuracy of the analytical results that we derive in the next sections, we  show in Fig.~\ref{psIV} results of numerical computations for matrices of the form \eqref{pseudo_map} for different values of the parameter $a$. Agreement is remarkable for all parameter values.

%==============================================
\section{Joint distribution of eigenvalue spacings for model RS}\label{transfer_op}

We now calculate asymptotic expressions for the joint distribution of eigenvalue spacings for the Lax matrix ensemble corresponding to the RS model. As discussed in the previous section, eigenvalues are such that within an interval of length $a$ from any eigenvalue there are exactly $[a]$ other eigenvalues. We introduce the rescaled nearest-neighbour spacings
\begin{equation}
\label{nns}
\xi_k\equiv \frac{N}{2\pi}\left(\lambda_{\alpha+1}-\lambda_{\alpha}\right)\ .
\end{equation}
The answer strongly depends on the integer part of $a$. Therefore we consider different cases separately.

\subsection{$0<a<1$}
The simplest case corresponds to $0<a<1$. In this case the only restriction is that the distance between the nearest eigenvalues is 
larger than $a$, namely $\xi_k\geq a$. It is convenient to rewrite this restriction as follows. 
The joint probability density of having $N+1$ eigenvalues inside an interval of length $L$ is given by
\begin{equation}
p(\xi_1,\xi_2,\ldots,\xi_{N})=\frac{1}{Z_N(L)}\prod_{j=1}^{N}g(\xi_{j})\delta \left (L-\sum_{k=1}^N\xi_k\right )  
\label{joint_proba_a_less_1}
\end{equation}
where 
\begin{equation}
g(x)=\left \{\begin{array}{cl} 0&\mathrm{when}\;x<a\\1&\mathrm{otherwise}\end{array}\right.\ ,
\label{function_g}
\end{equation}
and $Z_N(L)$ is the normalisation constant
\begin{equation}
Z_N(L)=\int_0^{\infty}\mathrm{d}\xi_1 \ldots \int_0^{\infty}\mathrm{d}\xi_N \prod_{j=1}^{N}g(\xi_{j})\delta \left (L-\sum_{k=1}^N\xi_k\right)\ .
\end{equation}
We are interested in the joint probability distribution of $n$ consecutive spacings
\begin{equation}
\label{px1xn}
p(\xi_1,\ldots,\xi_n)=\int_0^{\infty}\mathrm{d}\xi_{n+1} \ldots \int_0^{\infty}\mathrm{d}\xi_N p(\xi_1,\xi_2,\ldots,\xi_{N})
\end{equation}
when $L,N\to\infty$ with mean level spacing $\Delta=L/N$ remaining constant. In the following we set 
$\Delta=1$. We shall proceed as it was done in \cite{GerlandEpjb}. The multiple integrals in 
\eqref{px1xn} are easily calculated by introducing the function
\begin{equation}
h_{n,N}(L)=\int_0^{\infty}\mathrm{d}\xi_{n+1} \ldots \int_0^{\infty}\mathrm{d}\xi_N \prod_{j=1}^{N}g(\xi_j)\delta \left (L-\sum_{k=1}^N\xi_k\right), 
\end{equation}
whose Laplace transform reads
\begin{equation}
g_{n,N}(t)=\lambda(t)^{N-n}\prod_{k=1}^{n}g(\xi_{k})e^{-t\xi_{k}}\, .
\end{equation}
Here
\begin{equation}
\lambda(t)=\int_0^{\infty}g(x)\mathrm{e}^{-tx}\mathrm{d}x=\frac{\mathrm{e}^{-ta}}{t}
\end{equation}
is the Laplace transform of $g(x)$. The inverse Laplace transform of $\lambda(t)^{N-n}$ is then
\begin{equation}
\label{lapinverse}
\frac{1}{2i\pi}\int_{c-i\infty}^{c+i\infty}\lambda(t)^{N-n}\exp(L t)\, \mathrm{d}t
=\frac{1}{2i\pi}\int_{c-i\infty}^{c+i\infty}\frac{1}{\lambda(t)^n}
\exp\left(N(\ln\lambda(t)+\Delta t)\right)\, \mathrm{d}t.
\end{equation}
The large-$N$ behaviour of \eqref{lapinverse} is obtained by saddle-point approximation. The value $c$ corresponds to the solution of the saddle-point equation 
\begin{equation}
1+\frac{\lambda^{\prime}(c)}{\lambda(c)}=0 
\end{equation}
(recall that $\Delta=1$). The solution reads
\begin{equation}
c=\frac{1}{1-a}\ .
\end{equation}
Similarly, the  inverse Laplace transform of $g_{n,N}(t)$ can be calculated by saddle-point approximation. Rather than calculating explicitly all prefactors coming from the integration, it is easier to observe that the large-$N$ behaviour of the normalisation factor $Z_N=h_{0,N}(L)$ is obtained similarly. One finally gets
\begin{equation}
\label{pxi1xin01}
p(\xi_1,\ldots,\xi_n)=\frac{\mathrm{e}^{n a/(1-a)}}{(1-a)^n}\prod_{j=1}^ng(\xi_j)
\mathrm{e}^{-\xi_j/(1-a)}\, .
\end{equation}

\subsection{$1<a<2$}
We now consider the case $1<a<2$. According to previous section for sufficiently 
large matrix size, any eigenvalue $x_k$ is such that there exists exactly $1$ eigenvalue in 
the interval $]x_k,x_k+2\pi a/N[$. In other words, the constraints on $x_k$ are that 
$x_{k+1}\in]x_k,x_k+2\pi a/N[$ and $x_{k+2}>x_k+2\pi a/N$. In terms of  the differences \eqref{nns}
between consecutive eigenvalues, the above restriction are equivalent to
\begin{equation}
0<\xi_{k+1}<a\;,\;\;a< \xi_{k+1}+\xi_k\ .
\end{equation}
Introduce the function $f(x)$  by 
\begin{equation}
f(x)=\left \{\begin{array}{cl} 1&\mathrm{when}\;0<x<a\\0&\mathrm{otherwise}\end{array}\right .
\label{function_f}
\end{equation}
and $g(x)$ as in \eqref{function_g}. Then the joint probability density of $N+1$ eigenvalues inside an interval
of length $L$ is given by the following expression 
\begin{equation}
p(\xi_1,\xi_2,\ldots,\xi_{N})=\frac{1}{Z_N(L)}\prod_{j=1}^{N}f(\xi_{j})g(\xi_{j}+\xi_{j+1})\delta \left (L-\sum_{k=1}^N\xi_k\right )  
\label{joint_proba}
\end{equation}
where $Z_N(L)$ is the normalisation constant
\begin{equation}
Z_N(L)=\int_0^{\infty}\mathrm{d}\xi_1 \ldots \int_0^{\infty}\mathrm{d}\xi_N \prod_{j=1}^{N}f(\xi_{j})g(\xi_{j}+\xi_{j+1})\delta \left (L-\sum_{k=1}^N\xi_k\right )\ .
\end{equation}
The large-$N$ behaviour of the joint probability distribution of $n$ consecutive spacings
\eqref{px1xn} can be then obtained as above, following \cite{GerlandEpjb}. We review the main 
steps of the procedure. Introducing the function
\begin{equation}
h_{n,N}(L)=\int_0^{\infty}\mathrm{d}\xi_{n+1} \ldots \int_0^{\infty}\mathrm{d}\xi_N \prod_{j=1}^{N}f(\xi_{j})g(\xi_{j}+\xi_{j+1})\,\delta \left (L-\sum_{k=1}^N\xi_k\right), 
\end{equation}
its Laplace transform reads
\begin{equation}
g_{n,N}(t)=\int_0^{\infty}\mathrm{d}\xi_{n+1} \ldots \int_0^{\infty}\mathrm{d}\xi_N \prod_{j=1}^{N}\mathrm{e}^{-t\xi_j}f(\xi_{j})g(\xi_{j}+\xi_{j+1})\ .
\label{trace}
\end{equation}
This quantity can be seen as a product of transfer operators
\begin{eqnarray}
\label{gnN}
g_{n,N}(t)&=&K_t(\xi_1,\xi_2)K_t(\xi_2,\xi_3)\ldots K_t(\xi_{n-1},\xi_n)\\
&\times&\int_0^{\infty}\mathrm{d}\xi_{n+1} \ldots \int_0^{\infty}\mathrm{d}\xi_N\,K_t(\xi_n,\xi_{n+1})\ldots K_t(\xi_{N-1},\xi_N)K_t(\xi_N,\xi_1)\, ,
\end{eqnarray}
where the transfer operator, $K_t(\xi,\xi^{\prime})$ is defined as
\begin{equation}
K_t(\xi,\xi^{\prime})=f(\xi)g(\xi+\xi^{\prime})f(\xi^{\prime})\,\mathrm{e}^{-t(\xi+\xi^{\prime})/2}\ .
\label{transfer}
\end{equation}
This is a real symmetric operator. Its real eigenvalues, $\lambda_j(t)$, and eigenfunctions, $\phi_j(t;\xi)$, verify
\begin{equation}
\label{ev_equation}
\int_0^{\infty}K_t(\xi,\xi^{\prime})\phi_j(t;\xi^{\prime})\mathrm{d}\xi^{\prime}=\lambda_j(t)\phi_j(t;\xi)\ .
\end{equation}
As this operator is real symmetric its eigenfunctions can be chosen to be orthonormal
\begin{equation}
\label{orthophi}
\int_0^{\infty}\phi_j(t;\xi)\phi_k(t;\xi)\mathrm{d}\xi=\delta_{jk}\ .
\end{equation}
The  transfer operator can be expanded over the basis of eigenfunctions as
\begin{equation}
\label{expansionkt}
K_t(\xi,\xi^{\prime})=\sum_j\lambda_j(t)\phi_j(t;\xi)\phi_j(t;\xi').
\end{equation}
In the large-$N$ limit the dominant contribution to \eqref{gnN} is given by the 
largest eigenvalue $\lambda_0(t)$. Then using the orthogonality property 
\eqref{orthophi} of the $\phi_j$ we get
\begin{equation}
g_{n,N}(t)\underset{N\to \infty}{\sim}\prod_{k=1}^{n-1}K_t(\xi_k,\xi_{k+1})\lambda_0(t)^{N-n+1}\phi_0(t;\xi_n)\phi_0(t;\xi_1)\ .
\end{equation}  
The large-$N$ behaviour of $h_{n,N}(L)$ is obtained by performing the inverse Laplace transform
of $\lambda_0(t)^{N-n+1}$. As in \eqref{lapinverse} the leading term is obtained by 
saddle-point approximation; here the saddle-point is such that
\begin{equation}
\label{point_c}
1+\frac{\lambda_0^{\prime}(c)}{\lambda_0(c)}=0 
\end{equation}
(again we take $\Delta=1$). Calculating in a similar way the large-$N$ behaviour of 
the normalisation factor $Z_N=h_{0,N}(L)$, one finally gets
\begin{equation}
p(\xi_1, \xi_2 ,\ldots, \xi_n)=\frac{1}{\lambda_0^{n-1}(c)}\phi_0(c;\xi_1)K_c(\xi_1,\xi_2)K_c(\xi_2,\xi_3)\ldots K_c(\xi_{n-1},\xi_n)\phi_0(c;\xi_n)\ .
\label{pxi1xin12}
\end{equation}  

\subsection{$m<a<m+1$}

The general case $m<a<m+1$ can be treated in a similar way. 
In this case each interval $]x_k,x_k+2\pi a/N[$ contains exactly $m$ eigenvalues $x_{k+1},\ldots,x_{k+m}$.
In terms of the rescaled differences \eqref{nns} between consecutive eigenvalues this 
condition is equivalent to the following two inequalities
\begin{eqnarray}
&&0<\xi_k+\xi_{k+1}+\ldots+\xi_{k+m-1}<a\\
&&a<\xi_k+\xi_{k+1}+\ldots+\xi_{k+m}\ .
\end{eqnarray}
In analogy with Eq.~\eqref{joint_proba}, the joint probability of eigenvalues spacings thus reads
\begin{equation}
p(\xi_1,\xi_2,\ldots,\xi_{N})=\frac{1}{Z_N(L)}\prod_{j=1}^{N}f(\xi_{j}+\ldots+\xi_{j+m-1})g(\xi_{j}+\ldots +\xi_{j+m})\delta \left (L-\sum_{k=1}^N\xi_k\right )  
\label{joint_general}
\end{equation}
with $f$ and $g$ defined by \eqref{function_f} and \eqref{function_g}. 
The large-$N$ behaviour is calculated as above by introducing a transfer operator, which 
in this case depends on two sets of variables
$\boldsymbol\xi=(\xi_1,\ldots,\xi_m$) and $\boldsymbol\xi^{\prime}=(\xi_1^{\prime},\ldots,\xi_m^{\prime})$ shifted by one unit i.e. $\xi_2=\xi_1^{\prime}$, $\xi_3=\xi_2^{\prime}$, $\ldots$, $\xi_m=\xi_{m-1}^{\prime}$ (see e.g. \cite{GerlandEpjb}). The explicit form of the transfer operator is the following  
\begin{eqnarray}
\label{transfer_m}
&&K(\boldsymbol\xi,\boldsymbol\xi^{\prime})=\delta (\xi_2-\xi_1^{\prime})\ldots \delta (\xi_m-\xi_{m-1}^{\prime})\nonumber\\
&&\times \mathrm{e}^{-t\xi_1/2}f(\xi_1+\ldots+\xi_m)g(\xi_1+\ldots +\xi_m+\xi_m^{\prime})f(\xi_1^{\prime}+\ldots+\xi_m^{\prime})\mathrm{e}^{-t\xi_m^{\prime}/2}\ .
\end{eqnarray}
The eigenvalue equation 
\begin{equation}
\int K(\boldsymbol\xi,\boldsymbol\xi^{\prime})\phi(\boldsymbol\xi^{\prime})\mathrm{d}\boldsymbol\xi^{\prime}=\lambda \phi(\boldsymbol\xi)
\end{equation}
reduces to a one-dimensional equation because of the $\delta$-functions appearing in the definition 
of the transfer operator. This equation can be written in the form 
\begin{equation}
\mathrm{e}^{-t\xi_1/2}\int_0^{\infty}\mathrm{e}^{-tz/2}g(\xi_1+\xi_2+\ldots+\xi_m+z)\phi(t;\xi_2,\ldots,\xi_m,z)\mathrm{d}z=
\lambda(t) \phi(t;\xi_1,\ldots,\xi_m)\ .
\label{eigen_equation}
\end{equation}
Here it is implicitly assumed that all variables $\xi_j>0$ and  
\begin{equation}
\phi(t;\xi_1,\ldots,\xi_m)=0\;\;\mathrm{when}\;\; \xi_1+\ldots+\xi_m>a\ .
\end{equation}
As above the largest eigenvalue, $\lambda_0(t)$ as well as the corresponding eigenfunction 
$\phi_0(t;\boldsymbol\xi)$, calculated at point $t=c$ obeying the same saddle-point 
condition Eq.~\eqref{point_c}, determine all correlation functions in the limit of large $N$.
The joint probability of $n$ consecutive spacings takes a different form for $n\leq m$ and $n>m$. 
For $n\leq m$
\begin{equation}
p(\xi_1,\ldots,\xi_n)=\int_0^a\mathrm{d}\xi_{n+1}\ldots \int_0^a\mathrm{d}\xi_{m} \phi_0(c;\xi_1,\ldots,\xi_m) \phi_0(c;\xi_m,\ldots,\xi_1), 
\end{equation} 
while for $n>m$
\begin{eqnarray}
\label{pxi1xinm}
&&p(\xi_1,\ldots,\xi_n)=\lambda_0(c)^{-n+m}\phi_0(c;\xi_n,\ldots,\xi_{n-m})\phi_0(c;\xi_1,\ldots,\xi_m)\mathrm{e}^{-c\sum_{s=1}^n\xi_s}\nonumber\\
&&\times \prod_{j=1}^{n-m+1}f(\xi_j+\ldots +\xi_{j+m-1})\prod_{j=1}^{n-m}g(\xi_j+\ldots+\xi_{j+m}). 
\end{eqnarray} 

%==========================================================================
\section{Nearest-neighbour spacing distributions for model RS}\label{correlation_functions}

In the previous section we have derived expressions for the joint distribution of eigenvalue spacings
$p(\xi_1,\ldots,\xi_n)$. From these expressions the $n$th nearest-neighbour spacing distribution can be calculated as
\begin{equation}
\label{pns}
P(n,s)=\int_0^{\infty}\mathrm{d}\xi_{1} \ldots \int_0^{\infty}\mathrm{d}\xi_n\, p(\xi_1,\ldots,\xi_n)\, 
\delta \left (s-\sum_{i=1}^n\xi_i\right)\ .
\end{equation}

\subsection{$0<a<1$}
For $0<a<1$ these integrals are easily calculable (e.g. by Laplace transform) and 
from the joint distribution Eq.~\eqref{pxi1xin01} we obtain
\begin{equation}
\label{pns01}
P(n,s)=\left \{\begin{array}{cr} \dfrac{\mathrm{e}^{(na-s)/(1-s)}}{(1-a)^n}\dfrac{(s-n a)^{n-1}}{(n-1)!},&s\geq na\\
0, &0<s<na \end{array} \right . .
\end{equation}
Comparison with numerical simulations is displayed at Fig.~\ref{psIV}.

\subsection{$1<a<2$}
When $1<a<2$ the joint distribution is given by Eq.~\eqref{pxi1xin12}.
What remains is to calculate the largest eigenvalue of
the transfer operator $K_t$, as well as its associated eigenfunction. As $K_t$
is a positive operator, the analog of the Perron-Frobenius theorem states 
that the eigenvector corresponding to the largest eigenvalue is positive. Orthogonality of 
the eigenfunctions implies that the converse is also true. Thus if one finds a positive 
eigenfunction then the corresponding eigenvalue is the largest one.
The eigenvalue equation~\eqref{ev_equation} is equivalent to
\begin{equation}
\mathrm{e}^{-t\xi /2}\int_{a-\xi}^a\mathrm{e}^{-t\xi^{\prime}/2}\phi(\xi^{\prime})\mathrm{d}\xi^{\prime}=\lambda \phi(\xi)\ .
\label{main}
\end{equation}
Let us look for solutions of Eq.~\eqref{main} positive on $[0,a]$
under the form $\phi(\xi)=\sinh \rho \xi$, with $\rho$ some unknown complex parameter.
Since Eq.~\eqref{main} should hold for all $\xi\in[0,a]$ we get the necessary condition
\begin{equation}
t=-2\rho\coth(\rho a).
\label{t_rho} 
\end{equation}
When $t<-2/a$, Eq.~\eqref{t_rho} admits two real solutions $\rho=\pm\rho_0$. Thus 
$\phi(\xi)=\sinh\rho_0\xi$ with $\rho_0>0$ solution of Eq.~\eqref{t_rho} is a positive solution 
of  Eq.~\eqref{main}. If $t>-2/a$,  Eq.~\eqref{t_rho} admits two pure imaginary solutions 
$\rho=\pm i\rho_0$,
thus $\phi(\xi)=\sin\rho_0\xi$  with $\rho_0>0$ and $i\rho_0$ solution of Eq.~\eqref{t_rho} 
is a positive solution of  Eq.~\eqref{main}. Finally if $t=-2/a$, $\rho=0$ is the unique solution to
Eq.~\eqref{t_rho}. In that case  $\phi(\xi)=\xi$ is a solution of Eq.~\eqref{main} which is 
positive on $[0,a]$. Thus for all $t$ we have a positive solution to Eq.~\eqref{main}. 
Properly normalised, this solution gives the eigenvector $\phi_0(t;\xi)$. The
corresponding eigenvalue is given by 
\begin{equation}
\lambda_0(t)=\frac{\mathrm{e}^{(\rho-t/2)a}}{\rho-t/2},
\label{lambda}
\end{equation}
with $\rho$ an implicit function of $t$.

The saddle-point $c$ is a solution of Eq.~\eqref{point_c}. For $\lambda_0$ given
by Eq.~\eqref{lambda} the condition becomes
\begin{equation}
1+2 a (2-a)\rho^2-\cosh 2\rho a+2\rho (a-1)\sinh 2\rho a=0
\end{equation}
and the saddle-point $c$ is obtained from $\rho$ through Eq.~\eqref{t_rho}. 
Equivalently, this condition can be expressed as
\begin{equation}
\label{eqaz}
a=\frac{2z^2-z\sinh 2z}{z^2+\sinh^2z-z\sinh 2z},\;\;\;\; z=\rho a\ .
\end{equation}
\begin{figure}[hbt]
\begin{center}
\includegraphics[width=.5\linewidth]{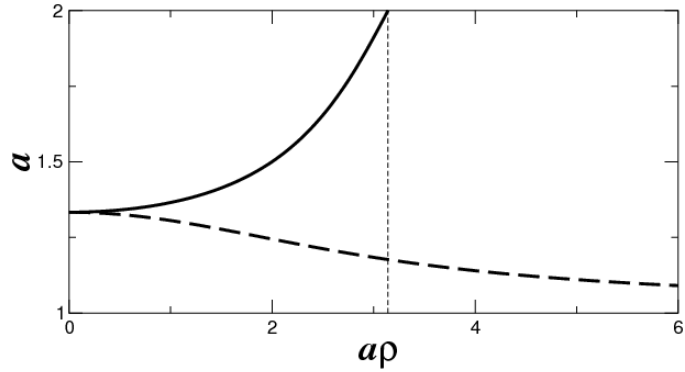}
\end{center}
\caption{(Color online). Graph of Eq.~\eqref{eqaz} for real (dashed) or pure imaginary (solid)
$z=\rho a$. Dashed vertical line indicates the abscissa equal $\pi$. }
\label{az}
\end{figure}

In Fig.~\ref{az} we plot $a$ as a function of $z=\rho a$. For $1<a<4/3$ Eq.~\eqref{eqaz} has a unique real 
solution $\rho_0>0$, and $\phi_0(\xi)=\sinh\rho_0\xi$ is a positive eigenfunction of the transfer 
operator. For $4/3<a<2$ Eq.~\eqref{eqaz} has a unique pure 
imaginary solution $i\rho_0$ with $\rho_0>0$. Furthermore in that latter case $\rho_0 a\in]0,\pi[$,
so that $\phi_0(\xi)=\sin\rho_0\xi$ is an eigenfunction of the transfer 
operator which is positive on $[0,a]$. At $a=4/3$ the unique solution is $\rho=0$ and $\phi_0(\xi)=\xi$
is a positive eigenfunction of the transfer operator. 
The $n$th nearest-neighbour spacing distribution can now be calculated from Eq.~\eqref{pxi1xin12}.

In the case $n=1$ it directly gives us the nearest-neighbour spacing 
distribution $P(s)=A^2\phi_0(s)^2$, where $A$ is the normalisation constant. 
It is nonzero only for $s\in[0,a]$, where it takes the following form
\begin{equation}
P(s)=\left \{\begin{array}{cl}A^2\sinh^2(\rho s)&\;\;\mathrm{when}\;\;1<a<4/3\\
\frac{81}{64}s^2&\;\;\mathrm{when}\;\;a=4/3\\
A^2\sin^2(\rho s)&\;\;\mathrm{when}\;\;4/3<a<2
\end{array}\right. .
\label{nearest}
\end{equation}
Constants $A$ and $\rho$ can be determined either by solving Eq.~\eqref{eqaz} and 
normalizing the eigenfunction $\phi_0(\xi)$, or equivalently by imposing the normalisation conditions \eqref{normalization}. The next-to-nearest distribution, $P(2,s)$ is non-zero only when $a<s<2a$ and within this 
interval it is given by  
\begin{equation}
P(2,s)=\frac{A^2}{\lambda_0(c)}\mathrm{e}^{-c s/2}\int_{s-a}^a\phi_0(\xi)\phi_0(s-\xi)\mathrm{d}\xi\ .
\end{equation}
In particular for $a=4/3$ all integrals can be calculated analytically and $P(2,s)$ has the form
\begin{equation}
P(2,s)=(-\frac{3}{2}+\frac{27}{16}s-\frac{81}{512}s^3)\mathrm{e}^{3s/4-1}\ .
\label{p2s}
\end{equation}
In a similar manner one can obtain the higher nearest-neighbour functions. 
For example, $P(3,s)$ is given by the formula
\begin{equation}
P(3,s)=\frac{A^2\mathrm{e}^{-c s/2}}{\lambda_0(c)^2}
\int_{0}^a\mathrm{d}\xi_1\phi_0(\xi_1)
\int_{a-\xi_1}^a\mathrm{d}\xi_2\mathrm{e}^{-c\xi_2/2}
\int_{a-\xi_2}^a\mathrm{d}\xi_3\phi_0(\xi_3)
\delta \left (s-\sum_{i=1}^n\xi_i\right).
\end{equation}
It is non-zero only when $a<s<3a$. In particular for $a=4/3$ we obtain
\begin{equation}
p(3,s)=\left \{\begin{array}{cc}(\frac{3}{4}-\frac{81}{32}s+\frac{81}{512}s^3)\mathrm{e}^{3s/4-1}+\frac{81}{64}s^2&\;\mathrm{when}\;4/3<s<8/3\\
(-\frac{9}{4}+\frac{27}{32}s-\frac{81}{512}s^3)\mathrm{e}^{3s/4-1}+9\mathrm{e}^{3s/2-4}&\;\mathrm{when}\;8/3<s<4
\end{array}\right . .
\label{p3s}
\end{equation} 
In Fig.~\ref{psIV} these formulas are compared with numerical simulations and show a remarkable agreement.

\subsection{$2<a<3$}

For $2<a<3$ the joint distribution is given by \eqref{pxi1xinm}.
The largest eigenvalue and corresponding eigenfunction of the transfer operator \eqref{transfer_m}
are solution of the eigenvalue  equation (\ref{eigen_equation}). For $m=2$ it takes the form
\begin{equation}
\mathrm{e}^{-t\xi_1/2}\int_{a-\xi_1-\xi_2}^{a-\xi_2}\mathrm{e}^{-t\xi_3/2}\phi(\xi_2,\xi_3)\mathrm{d}\xi_3=
\lambda \phi(\xi_1,\xi_2)\ .
\label{T_3}
\end{equation} 
Let us look for solutions of the form 
similar to Bethe Ansatz
\begin{eqnarray}
\phi(\xi_1,\xi_2)&=&\mathrm{e}^{\alpha \xi_1+\beta \xi_2}+\mathrm{e}^{-\beta \xi_1+(\alpha-\beta-\mu) \xi_2}+
\mathrm{e}^{(-\alpha+\beta+\mu) \xi_1-\alpha \xi_2}\nonumber\\
&&-\mathrm{e}^{-\beta \xi_1-\alpha \xi_2}-\mathrm{e}^{\alpha \xi_1+(\alpha-\beta-\mu) \xi_2}-
\mathrm{e}^{(-\alpha+\beta+\mu) \xi_1+\beta \xi_2}\ ,
\label{eigenfunction}
\end{eqnarray} 
where we have set $\mu=-t/2$. As Eq.~\eqref{T_3} has to be fulfilled for all $\xi_1,\xi_2$,
this function is a solution of (\ref{T_3}) if and only if the following conditions are valid
\begin{equation}
\frac{\mathrm{e}^{a(\mu+\beta)}}{\mu+\beta}=\frac{\mathrm{e}^{a(\mu-\alpha)}}{\mu-\alpha}=\frac{\mathrm{e}^{a(\alpha-\beta)}}{\alpha-\beta}=-\frac{\lambda}{a}\ .
\label{quantization}
\end{equation}
From the first equality in Eq.~\eqref{quantization} one can express $\mu$ as a function of 
$\alpha$ and $\beta$.
After inspection we found that the solutions of the above equations have the following form
\begin{equation}
\alpha a=\frac{1}{2}x_1+\mathrm{i}x_2\ ,\;\;\beta a=-\frac{1}{2}x_1+\mathrm{i}x_2\ ,\mu a=\frac{1}{2}x_1+x_3
\end{equation}
with real parameters $x_1$, $x_2$, and $x_3$. 
Under this substitution the eigenfunction (\ref{eigenfunction}) is transformed to
\begin{equation}
\phi(\xi_1,\xi_2)=\mathrm{e}^{x_1(\xi_1-\xi_2)/2a}
\left( \sin\frac{x_2(\xi_1+\xi_2)}{a}-\mathrm{e}^{(x_1-x_3)\xi_2/a}\sin\frac{x_2\xi_1}{a}
-\mathrm{e}^{-(x_1-x_3)\xi_1/a}\sin\frac{x_2\xi_2}{a}\right)
\label{phix1x2}
\end{equation}
where from \eqref{quantization} $x_1$, $x_2$, and $x_3$ must fulfill the following equalities
\begin{equation}
\frac{\mathrm{e}^{x_1}}{x_1}=\frac{\mathrm{e}^{x_3+\mathrm{i}x_2}}{x_3+\mathrm{i}x_2}=\frac{\mathrm{e}^{x_3-\mathrm{i}x_2}}{x_3-\mathrm{i}x_2}=-\frac{\lambda}{a}
\label{quant}
\end{equation}
and depend on time $t$ through the relation 
\begin{equation}
\label{time}
-\frac{t a}{2}=\frac{x_1}{2}+x_3\ .
\end{equation}
This implies that
\begin{equation}
\label{x3}
x_3=\frac{x_2}{\tan x_2}
\end{equation}
and, consequently, $x_1$ is related with $x_2$ as follows
\begin{equation}
\label{x1x2}
\frac{\mathrm{e}^{x_1}}{x_1}=\frac{\sin x_2}{x_2} \mathrm{e}^{x_2/\tan x_2}\ .
\end{equation}
The eigenfunction corresponding to the largest eigenvalue of the transfer operator is thus
given by \eqref{phix1x2} with $x_1$, $x_2$, and $x_3$ real parameters depending on $t$,
which must verify \eqref{x3} and 
\eqref{x1x2} and be such that $\phi(\xi_1,\xi_2)$ is a positive function over $[0,a]^2$.

The saddle-point condition is again given by Eq.~(\ref{point_c}). Using 
\eqref{quant}--\eqref{x1x2} we get a second relation between $x_1$ and $x_2$, namely
\begin{equation}
a=\frac{1}{1-1/x_1}+\dfrac{2-\sin(2x_2)/x_2}{1+\sin^2(x_2)/x_2^2-\sin(2x_2)/x_2}\ .
\label{eqn_a}
\end{equation}
Equations (\ref{x1x2}) and (\ref{eqn_a}) determine parameters $x_1$ and $x_2$ at a given 
$a$. To get a positive eigenfunction $\phi$ it is necessary to get the solutions in 
the intervals
\begin{equation}
x_1<0 ,\;\;\pi<x_2<2\pi\ .
\end{equation}
The knowledge of these parameters allows us to calculate the eigenfunction \eqref{phix1x2},
from which the nearest-neighbour distributions can be deduced through Eq.~\eqref{pxi1xinm}.
The first distributions read
\begin{eqnarray}
\label{p1s23}
P(s)&=&A\int_0^{a-s}\phi(s,y)\phi(y,s)\mathrm{d}y\ ,\\
P(2,s)&=&A\int_0^{s}\phi(s-y,y)\phi(y,s-y)\mathrm{d}y\ ,
\label{p2s23}
\end{eqnarray}
and
\begin{equation}
\label{p3s23}
P(3,s)=\frac{A}{\lambda}\int_{s-a}^{a}\mathrm{d}x \mathrm{e}^{\mu x} \int_{s-a}^{s-x}\mathrm{d}y\mathrm{e}^{\mu y}\phi(s-x-y,x)\phi(s-x-y,y)\mathrm{d}y\ ,
\end{equation}
with $A$ the normalisation constant
\begin{equation}
A=\left (\int_0^{a}\mathrm{d}x \left (\int_0^{a-x}\mathrm{d}y \phi(x,y)\phi(y,x)\right )\right )^{-1}\ .
\end{equation}
These analytical expressions perfectly agree with numerical simulations, as shown in Fig.~\ref{psIV}.

%==============================================
\section{Level compressibility for model RS}\label{compressibility}

The expressions for the joint distribution of eigenvalue spacings
$p(\xi_1,\ldots,\xi_n)$ obtained in section \ref{transfer_op} allow to derive
formulas for the level compressibility $\chi$, which characterises the asymptotic behaviour
of the number variance.

The number variance $\Sigma^2(L)$ is the average variance of the number of energy levels 
in an interval of length $L$. It is defined from the two-point correlation function
$R_2(s)=\sum_{n=1}^{\infty}P(n,s)$ as
\begin{equation}
\Sigma^2(L)=L-2\int_0^Lds\, (L-s)(1-R_2(s)).
\end{equation}
For systems with intermediate spectral statistics, $\Sigma^2(L)\sim\chi L$ for large $L$.
In order to obtain the large-$N$ behaviour of the level compressibility we calculate 
the Laplace transform of the two-point correlation function. It
has a series expansion of the form
\begin{equation}
\label{lapR2}
g_2(t)=\int_0^Lds\, R_2(s)\mathrm{e}^{-t s}=\frac{1}{t}+\frac{\chi-1}{2}+O(t)
\end{equation}
which allows us to obtain $\chi$ (see~\cite{GerlandEpjb} for more detail).

\subsection{Case $0<a<1$}

The $n$th nearest-neighbour spacing distributions are given by 
Eq.~\eqref{pns01}. Summation over $n$ gives the two-point correlation function, and its Laplace transform is readily obtained, yielding
\begin{equation}
g_2(t)=\frac{1}{\mathrm{e}^{a t}(1+t-a t)-1}.
\end{equation}
Small-$t$ expansion of $g_2(t)$ gives
\begin{equation}
\label{chi01}
\chi=(1-a)^2.
\end{equation}

\subsection{Case $1<a<2$}

The functions $P(n,s)$ are given by Eqs.~\eqref{pxi1xin12} and \eqref{pns}. Their Laplace transform reads
\begin{eqnarray}
g(n,t)&=&\frac{1}{\lambda_0^{n-1}(c)}
\int_0^{\infty}\!\!\mathrm{d}\xi_{1} \ldots \int_0^{\infty}\!\!\mathrm{d}\xi_n\nonumber\\
&\times & \phi_0(c;\xi_1)K_c(\xi_1,\xi_2)K_c(\xi_2,\xi_3)\ldots K_c(\xi_{n-1},\xi_n)\phi_0(c;\xi_n)
\mathrm{e}^{-t(\xi_{1}+\ldots+\xi_n)},
\label{gnt12}
\end{eqnarray}
where as in the previous sections $\lambda_0(c)$ is the largest eigenvalue 
of the transfer operator \eqref{transfer} and $\phi_0(c;\xi)$ its associated eigenfunction,
both taken at the saddle-point $c$.
Using the definition \eqref{transfer} of the transfer operator, we see that $g(n,t)$
can be rewritten 
\begin{equation}
g(n,t)=\frac{1}{\lambda_0^{n-1}(c)}\int_0^{\infty}\!\!\mathrm{d}\xi
\int_0^{\infty}\!\!\mathrm{d}\xi'\phi_0(c;\xi)\mathrm{e}^{-t\xi/2}K_{c+t}(\xi,\xi')^{n-1}
\phi_0(c;\xi')\mathrm{e}^{-t\xi'/2}\ .
\end{equation}
Replacing the transfer operator by its expansion \eqref{expansionkt} and summing over $n$
we get
\begin{equation}
g_2(t)=\sum_j\frac{\lambda_0(c)}{\lambda_0(c)-\lambda_j(c+t)}
\left(\int_0^{\infty}\!\!\mathrm{d}\xi\phi_0(c;\xi)\phi_j(c+t;\xi)\mathrm{e}^{-t\xi/2}\right)^2.
\end{equation}
One can check, using normalisation \eqref{orthophi} of the eigenfunctions and
the saddle-point condition \eqref{point_c}, that the leading-order term is
given by $g_2(t)\sim 1/t$. The next-order term can be simplified using
the normalisation of $\phi_0$. It yields
\begin{equation}
g_2(t)=\frac1t-\frac{\lambda_0''(c)}{2\lambda_0'(c)}-1+o(t^2),
\end{equation}
from which one gets
\begin{equation}
\label{chi_vs_lambda}
\chi=-1-\frac{\lambda_0''(c)}{\lambda_0'(c)}.
\end{equation}
Here $\lambda_0(t)$ is given by \eqref{lambda} (with $\rho$ depending on $t$ through
\eqref{t_rho}), and $c$ is given by condition \eqref{point_c}. After calculation,
$\chi$ can be expressed as a function of $\rho$ at the saddle-point. We get
\begin{equation}
\label{chi12}
\chi=\left(\frac{a^2}{4}-\frac{4a(1-a)z^2+a^2\sinh^2z}{(2z-\sinh 2z)^2}
\sinh^2z\right)\frac{\sinh^2z}{z^2},\,\,\,\,\,\, z=\rho a,
\end{equation}
with $\rho$ the real positive solution $\rho_0$ of \eqref{eqaz} for $1<a<4/3$ or the 
pure imaginary solution $i\rho_0$ of \eqref{eqaz} for $4/3<a<2$. For $a=4/3$, the
limit $\rho\to 0$ in \eqref{chi12} gives $\chi=4/9$. 

\subsection{Case $2<a<3$}

\begin{figure}[ht]
\begin{center}
\includegraphics[width=.65\linewidth]{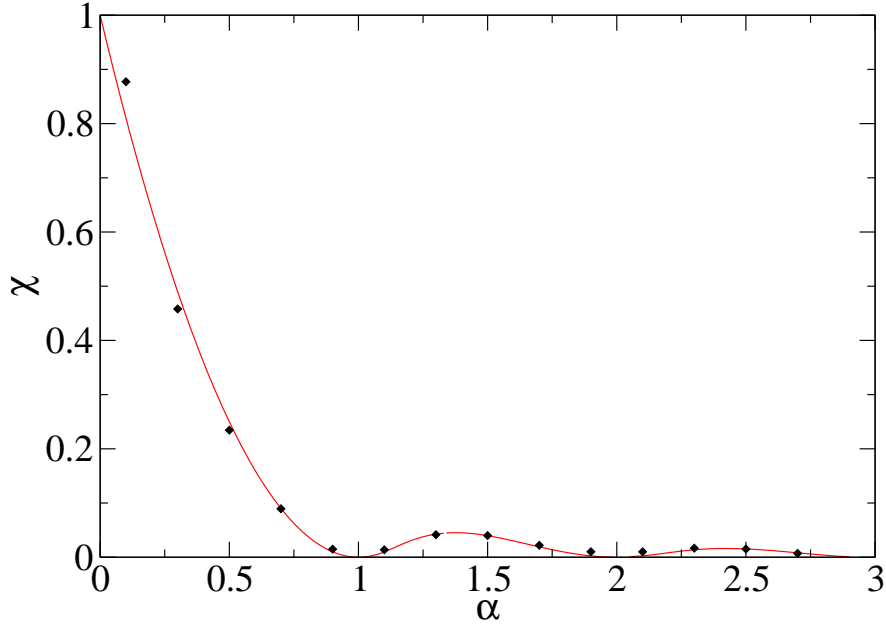}
\end{center} 
\caption{Level compressibility $\chi$. Black diamonds are the numerical values extracted from a cubic fit of $\Sigma^2(L)$ over the range $L\in[0,80]$, with $\Sigma^2(L)$ the variance of the number of levels in an interval of length $L$ averaged over 50 windows of length $L$, calculated from the unfolded spectrum with mean level spacing $\Delta=1$, for 10000 realisations of the random matrix with matrix size $N=256$. Red curve is the theoretical prediction \eqref{chi01}, \eqref{chi12} and \eqref{chi23}. \label{chi_ruij}}
\end{figure}

As in the previous case $\chi$ is given by \eqref{chi_vs_lambda} with $\lambda_0(t)$ 
given by \eqref{quant}, with $x_1$, $x_2$, $x_3$ and $t$ related through 
\eqref{time}--\eqref{x1x2}. From \eqref{time}--\eqref{x3}, parameter $x_1$ 
can be expressed as 
\begin{equation}
\label{x1x2t}
x_1=-2\frac{x_2}{\tan x_2}-a t.
\end{equation}
Differentiating both \eqref{x1x2} and \eqref{x1x2t} with respect to time we obtain 
$\mathrm{d}x_1/\mathrm{d}t$ and $\mathrm{d}x_2/\mathrm{d}t$ as a function of $x_1$ and $x_2$, and then similarly $\mathrm{d}^2x_1/\mathrm{d}t^2$ and $\mathrm{d}^2x_2/\mathrm{d}t^2$. Using \eqref{quant},
the saddle-point condition \eqref{point_c} can be rewritten
\begin{equation}
1+\frac{\mathrm{d}x_1}{\mathrm{d}t}\left(1-\frac{1}{x_1}\right)=0,
\end{equation}
and from \eqref{chi_vs_lambda} $\chi$ can then be expressed as
\begin{equation}
\chi=\frac{1}{(1-x_1)^2}+\frac{\mathrm{d}^2x_1}{\mathrm{d}t^2}\left(1-\frac{1}{x_1}\right).
\end{equation}
Using the expression obtained $\mathrm{d}^2x_1/\mathrm{d}t^2$ we finally obtain
$\chi$ as a function of $x_1$ and $x_2$, with $x_1$, $x_2$ obtained as solution of
\eqref{x1x2}--\eqref{eqn_a}. Inverting \eqref{eqn_a} we get
\begin{equation}
\label{x1x2bis}
x_1=\frac{a\sin^2x_2+(a-2)x_2^2+(1-a)x_2\sin 2x_2}{(a-1)\sin^2x_2+(a-3)x_2^2+(2-a)x_2\sin 2x_2}\ .
\end{equation}
After some manipulation $\chi$ simplifies to
\begin{eqnarray}
\label{chi23}
\chi&=&\frac{1}{a(\sin^2x_2+x_2^2-x_2\sin 2x_2)^2}\left[
(a-3)^2(a-2)x_2^4\right.\\
&-&(a-3)(a-1)(2a-5)x_2^3\sin 2x_2+2(a-2)((\cos 2x_2+2)(a-1)(a-2)-3)x_2^2\sin^2x_2\nonumber\\
&-&\left. 2a(a-2)(2a-3)x_2\cos x_2\sin^3 x_2+a(a-1)^2\sin^4x_2\right]\ .\nonumber
\end{eqnarray}
Figure \ref{chi_ruij} is a plot of the level compressibility $\chi$. The theoretical prediction obtained from \eqref{chi01}, \eqref{chi12} and \eqref{chi23} agrees with numerical data.

\subsection{Asymptotics in the vicinity of integer $a$}

For integer $a$ the spectrum is rigid and thus the level compressibility is expected to take the value 0. Here we consider the first-order expansion of $\chi$ in the vicinity of integer $a$. We will show that at lowest order the expansion of $\chi$ around $a=n$ is given by $\chi\simeq (1-a)^2/n^2$.

We first consider the expansion around $a=1$. Let $a=1+\epsilon$. For $a<1$ we have $\chi=(1-a)^2=\epsilon^2$, thus expansion is trivial. For $a>1$ $\chi$ is given by \eqref{chi12} with $a$ and $z$ related by \eqref{eqaz}. At $a=1$ the solution of \eqref{eqaz} is $z=\infty$. An asymptotic expansion of \eqref{chi12} and \eqref{eqaz} yields 
\begin{equation}
a=\frac{2z}{2z-1}+F(z)\mathrm{e}^{-2z}+o(\mathrm{e}^{-2z})
\end{equation}
where $F(z)$ is some rational fraction in $z$. Thus at first order 
\begin{equation}
\label{zeps}
z=\frac12\left(1+\frac{1}{\epsilon}\right).
\end{equation}
Expanding $\chi$ for large $z$ to lowest order gives
\begin{equation}
\chi=a(a-1)+a^2\left(\frac{1}{4z^2}-\frac{1}{2z}\right)+G(z)\mathrm{e}^{-2z}+o(\mathrm{e}^{-2z})
\end{equation}
with $G(z)$ is some rational fraction in $z$. Using \eqref{zeps} one gets $\chi\simeq\epsilon^2=(1-a)^2$.

Suppose now that $a=2+\epsilon$. For $a<2$ $\chi$ is given by \eqref{chi12}, with 
$\rho=i\rho_0$ solution of \eqref{eqaz}. Equivalently, $\chi$ is given by
\begin{equation}
\label{chi2}
\chi=\left(\frac{a^2}{4}+\frac{4a(1-a)z^2+a^2\sin^2z}{(2z-\sin 2z)^2}
\sin^2z\right)\frac{\sin^2z}{z^2},
\end{equation}
with $z$ the real positive solution of 
\begin{equation}
\label{eqaz2}
a=\frac{2z^2-z\sin 2z}{z^2-2z\sin 2z+\sin^2z}.
\end{equation}
At point $a=2$ the solution is $z=\pi$. Expanding both sides of \eqref{eqaz2} at lowest order in $\epsilon$ with $z=\pi+z_1 \epsilon$ we get $z1=\pi/2$. Inserting this expansion for $z$ in \eqref{chi2} we obtain $\chi=\epsilon^2/4+o(\epsilon^3)$.

For $a>2$ $\chi$ is given by \eqref{chi23}, with $x_1$ and $x_2$ specified by \eqref{x1x2}--\eqref{eqn_a}. At $a=2$, we have
\begin{equation}
\chi=\frac{2\sin^4 x_2-x_2^3\sin 2x_2}{2(x_2^2+\sin^2x_2-x_2\sin 2x_2)}
\end{equation}
which vanishes for $x_2=\pi$. For $x_2=\pi+t_1\epsilon+t_2\epsilon^2$ we have the expansion
\begin{equation}
\label{chiexp}
\chi=\left(\frac12-\frac{t_1}{\pi}\right)\epsilon+\left(-\frac54+\frac{9t_1}{2\pi}-\frac{3t_1^2}{\pi^2}-\frac{t_2}{\pi}\right)\epsilon^2+o(\epsilon^3).
\end{equation}
Equation \eqref{x1x2} is equivalent to
\begin{equation}
\exp\left(x_1-\frac{x_2}{\tan x_2}\right)=\frac{x_1 \sin x_2}{x_2}
\end{equation}
and the small-$\epsilon$ expansion of both members of this equation reads
\begin{eqnarray}
x_1-\frac{x_2}{\tan x_2}&=&-\frac{\pi}{\epsilon t_1}+\frac{\pi t_2}{t_1}-1+o(1),\\
\frac{x_1 \sin x_2}{x_2}&=&-\frac{\pi t_1-2t_1^2}{\pi^2}\epsilon^2+\frac{\pi^2 t_1-5 \pi t_1^2 + 6 t_1^3 
+\pi^2 t_2-4 \pi t_1 t_2}{\pi^3}\epsilon^3+o(\epsilon^4)
\label{fracx1x2}
\end{eqnarray}
(we use \eqref{x1x2bis} to obtain the expansion of $x_1$).
This implies that the two first terms in the expansion \eqref{fracx1x2} must vanish, 
thus $t_1=\pi/2$ and $t_2=0$. Putting these values into \eqref{chiexp} gives 
$\chi=\epsilon^2/4+o(\epsilon^3)$.

The same result can be obtained from \eqref{chi23} and \eqref{x1x2}--\eqref{eqn_a} for $a<3$. At $a=3$ again $x_2$ takes the value $\pi$, and an expansion $x_2=\pi+t_1\epsilon$ gives $t_1=\pi/3$, whence 
$\chi=\epsilon^2/9+o(\epsilon^3)$.

%===============================================
\section{Conclusion}

In this paper we construct new random matrix ensembles with unusual properties. Random matrices from these ensembles are Lax matrices of $N$-body integrable classical systems with a certain measure of momenta and coordinates. Though such matrices are not invariant over rotation of the basis (as usual random matrix ensembles) the joint distribution of their eigenvalues can be calculated analytically. Four different models are considered in detail. Three of them correspond to rational, hyperbolic, and trigonometric Calogero-Moser models. The fourth is related to the trigonometric Ruijsenaars-Schneider model. For the trigonometric Calogero-Moser model and the  Ruijsenaars-Schneider model spectral correlation functions are calculated explicitly. For rational and hyperbolic  Calogero-Moser models Wigner-type surmises are proposed. Our formulas are in a good agreement with results of direct numerical calculations.    

%=======================================================================================
\appendix

\section{Hamilton-Jacobi equations}
\label{HJ}

In this appendix we check that the action-angle variables $\lambda_{\alpha}$ and $\phi_{\alpha}$ for model CM$_r$ verify Hamilton-Jacobi equations by calculating their time derivative. We use the fact that for the Lax pair $(L,M)$ the matrix $M$ can be seen as a time derivative operator for the eigenfunctions of the matrix $L$. Namely, if $(u_k)_{1\leq k \leq N}$ is a normalised eigenvector of $L$, then
\begin{equation}
\dot{u}_k=\sum_r M_{kr}u_r\;\;\;\; \textrm{and }\; \dot{u}_k^*=-\sum_r u_r^*M_{rk}\
\label{dot_M}
\end{equation}
(here $*$ denotes complex conjugation). For model CM$_r$, the Lax matrix $M$ is given by 
\begin{equation}
M_{kr}=-ig\,\delta_{kr}\sum_{j\neq k}\frac{1}{(q_k-q_j)^2} +i g (1-\delta_{kr})\frac{1}{(q_k-q_r)^2}\,. 
\label{matrix_M}
\end{equation}
One can easily check that for $1\leq k,r\leq N$ one has
\begin{equation}
\label{linkLMInr}
p_r\delta_{kr}+M_{kr}(q_k-q_r)=L_{kr}\ .
\end{equation}
Deriving \eqref{def_phi} with respect to time, using the definition of $Q$, yields
\begin{equation}
\dot{\phi}_{\alpha}=\sum_k [ \dot{u}_k^{*}(\alpha)q_k u_k(\alpha)+
u_k^*(\alpha)\dot{q}_k u_k(\alpha)+u_k^*(\alpha)q_k \dot{u}_k(\alpha)] \ .
\end{equation}
From Hamilton-Jacobi equations $\dot{q}_k=p_k$. Using \eqref{dot_M}, we get that the time derivative of $\phi_{\alpha}$ is given by
\begin{eqnarray}
\dot{\phi}_{\alpha}= \sum_{k,r} u_k^*(\alpha)\left[p_k\delta_{kr}+M_{kr}(q_k-q_r)\right]u_r(\alpha)
&=&\sum_{k,r}u_k^{*}(\alpha) L_{kr} u_r(\alpha)=\lambda_{\alpha}\ .
\end{eqnarray}
The time derivative of  $\lambda_{\alpha}$ is easily obtained from \eqref{dot_L},\eqref{dot_M} and \eqref{def_Lu}, yielding $\dot{\lambda}_{\alpha}=0$. This shows that the $\lambda_{\alpha}$ and $\phi_{\alpha}$ verify Hamilton-Jacobi equations.

\section{Identities}\label{app_A} 

The purpose of the Appendix is to give, for completeness, the proofs of certain often used formulas.

Let  coefficients $b_m$ obey the following system of linear equations for all $n=1,\ldots, N$  with known $x_m$ and $y_m$
\begin{equation}
 \sum_{m=1}^N\frac{b_m}{x_m-y_n}=1\;.
\label{eqn}
\end{equation}
Then $b_m$ for all $m=1,\ldots, N$ are expressed through $x_m$ and $y_m$ by using e.g. the Cauchy determinants 
\begin{equation}
b_m=\frac{\prod_n(x_m-y_n)}{\prod_{s\neq m}(x_m-x_s)}\ . 
\label{bm}
\end{equation}
The following identities are also useful. For all $l=1,\ldots, N$ one has 
\begin{equation}
 \sum_{m=1}^N\frac{b_m}{(x_m-y_l)^2}=-\frac{\prod_{n\neq l}(y_l-y_n)}{\prod_{s}(y_l-x_s)}\ ,
\label{sum_bm2}
\end{equation}
\begin{equation}
\sum_{m=1}^N b_m=\sum_{m=1}^N(x_m-y_m), 
\label{sum_bm}
\end{equation}
and
\begin{equation}
\sum_{m=1}^N \frac{b_m}{x_m}=1-\prod_{n}\frac{y_n}{x_n}\ .
\label{sum_norm}
\end{equation}
A simple way to check (\ref{bm}) is to consider the function
\begin{equation}
f_n(x)=\frac{\prod_{r\neq n}(x-y_r)}{\prod_{s}(x-x_s)}=\frac{\prod_{r}(x-y_r)}{(x-y_n)\prod_{s}(x-x_s)}\;.
\end{equation}
Asymptotically this function decreases as $1/x$ when $x\to\infty$, so that the integral over a large contour encircling all poles equals 1. Rewriting this integral as the sum over all finite poles gives
\begin{equation}
1=\sum_m\frac{\prod_r(x_m-y_r)}{(x_m-y_n)\prod_{s\neq m}(x_m-x_s)}
\end{equation}
which proves (\ref{bm}).
 
The equality (\ref{sum_bm2})  can be obtained by the integration of the function
\begin{equation}
 \bar{f}_l(x)=\frac{\prod_n(x-y_n)}{(x-y_l)^2\prod_{n}(x-x_n)}
\end{equation}
over a contour which includes all poles. As this function decreases as $1/x^2$ when $x\to\infty$ the integral equals zero. Taking the sum over  poles at $x=x_n$ with all $n=1,\ldots, N$ and at $x=y_l$ one verifies (\ref{sum_bm2}). 

To get \eqref{sum_bm} one has to integrate the function
\begin{equation}
 \tilde{f}(x)=\prod_{n=1}^N\frac{x-y_n}{x-x_n}
\end{equation}
over the large contour and compare the residues at infinity and at $x=x_n$.

Let us now consider the function
\begin{equation}
\hat{f}(x)=\frac{\prod_n(x-y_n)}{x\prod_{n}(x-x_n)}\ .
\end{equation}
It decreases as $1/x$ when $x\to\infty$ and has poles at $x=0$ and $x=x_m$ with $m=1,\ldots, N$. Integrating it over a contour encircling all poles one obtains (\ref{sum_norm}).

%============================================================


\begin{thebibliography}{99}
\bibitem{GuhMueWei98}T.~Guhr, A.~M{\"u}ller-Groeling, H.~A.~Weidenm{\"u}ller, \textit{Random Matrix Theories in Quantum Physics: Common Concepts}, Phys. Rep. {\bf 299}, 189 (1998).
\bibitem{BogGiaSch84} O. Bohigas, M.-J. Giannoni and C. Schmit, \textit{Characterization of chaotic quantum spectra and universality of level fluctuation laws}, Phys. Rev. Lett.  {\bf 52}, 1 (1984).
\bibitem{BerTab77} M.~V.~Berry and M.~Tabor, \textit{Level clustering in the regular spectrum},  
Proc. Roy. Soc. A {\bf 356}, 375 (1977). 
\bibitem{Meh90} M.~L.~Mehta, {\it Random Matrix Theory}, Springer, New York (1990).
\bibitem{zirnbauer} A.~Altland and M.~R.~Zirnbauer, \textit{Novel symmetry classes in mesoscopic \\normal-superconducting hybrid structures}, Phys. Rev. B \textbf{55}, 1142 (1997). 
\bibitem{abul} A.~Y.~Abul-Magd and M.~H.~Simbel, \textit{Wigner surmise for high-order level spacing distributions of chaotic systems}, Phys. Rev. E \textbf{60}, 5371 (1999).
\bibitem{anderson} P. W. Anderson,  \textit{Absence of Diffusion in Certain Random Lattices}, Phys. Rev. \textbf{109}, 1492 (1958). 
\bibitem{mit} B. I. Shklovskii, B. Shapiro, B. R. Sears, P. Lambrianides, and H. B. Shore, \textit{Statistics of spectra of disordered systems near the metal-insulator transition}, Phys. Rev. B \textbf{47}, 11487 (1993). 
\bibitem{girdif} E.~B.~Bogomolny, O.~Giraud and C.~Schmit, \textit{Nearest-neighbor distribution for singular billiards}, Phys. Rev. E {\bf 65}, 056214 (2002).
\bibitem{giraud} O.~Giraud, J.~Marklof and S.~O'Keefe, \textit{Intermediate statistics in quantum maps},  J. Phys. A
{\bf 37}, L303 (2004).
\bibitem{huckenstein} B.~Huckestein, \textit{Scaling theory of the integer quantum Hall effect}, Rev. Mod. Phys. {\bf 67}, 357 (1995).
\bibitem{Gerland} E. Bogomolny, U. Gerland, and C. Schmit, \textit{Models of intermediate spectral statistics}, Phys. Rev. E \textbf{59}, R1315 (1999) 
\bibitem{PRBM} A.~D.~Mirlin, Y.~V.~Fyodorov, F.-M.~Dittes, J.~Quezada, and T.~H.~Seligman, 
\textit{Transition from localized to extended eigenstates in the ensemble of power-law random banded matrices}, Phys. Rev. E {\bf 54}, 3221 (1996). 
\bibitem{kravtsov} V.~E.~Kravtsov and K.~A.~Muttalib,  \textit{New Class of Random Matrix Ensembles with Multifractal Eigenvectors}, Phys. Rev. Lett. {\bf 79}, 1913 (1997).
\bibitem{prl} E. Bogomolny, O. Giraud, and C. Schmit, \textit{Random Matrix Ensembles Associated with Lax Matrices}, 
Phys. Rev. Lett. {\bf 103},  054103 (2009).
\bibitem{fractal} E. Bogomolny and O.~Giraud, \textit{Perturbation approach to fractal dimensions for certain critical random matrix ensembles}, in preparation (2011).
\bibitem{lax} P. Lax, \textit{Integrals of nonlinear equations of evolution and solitary waves}, Comm. Pure Applied Math. \textbf{21}, 467 (1968).
\bibitem{I} S.N.M. Ruijsenaars, \textit{Action-angle maps and scattering theory for some finite-dimensional integrable systems I. The pure soliton case}, Commun. Math. Phys. \textbf{115}, 127 (1988).
\bibitem{II} S.N.M. Ruijsenaars, \textit{Action-angle maps and scattering theory for some finite-dimensional integrable systems II. Solitons, antisolitons, and their bound states}, PubL. RIMS, Kyoto Univ. \textbf{30}, 865 (1994).
\bibitem{III} S. Ruijsenaars, \textit{Action-angle maps and scattering theory for some finite-dimensional integrable systems, III. Sutherland type systems and their duals}, PubL. RIMS, Kyoto Univ. \textbf{31}, 247 (1995).
\bibitem{calogero} F. Calogero, \textit{Solution of the one-dimensional N-body problem with quadratic
and/or inversely quadratic pair potentials}, J. Math. Phys. \textbf{12}, 419 (1971); Erratum:
J. Math. Phys. \textbf{37}, 3646 (1996).
\bibitem{moser} J. Moser, \textit{Three integrable Hamiltonian systems connected with isospectral deformations},
Advances in Math., \textbf{16}, 197 (1975).
\bibitem{rs} S.N.M. Ruijsenaars and H. Schneider, \textit{A new class of integrable systems and its relation to solitons}. Ann. Phys. (NY) \textbf{170}, 370 (1986).
\bibitem{perelomov} M.A. Olshanetsky and A.M. Perelomov, \textit{Classical integrable finite-dimensional systems related to Lie algebras}, Phys. Rep. \textbf{71}, 313 (1981).
\bibitem{hoker} E. D'Hoker and D.H. Phong, \textit{Lax pairs and spectral curves for Calogero-Moser and spin Calogero-Moser systems}, arXiv:hep-th/9903002, (1999), Regul. Chaotic Dyn. \textbf{3}, 27 (1998).
\bibitem{krichever} I.M. Krichever, \textit{Elliptic solutions of the Kadomtsev-Petviashvili equation and integrable systems of particles}, Func. Anal. Appl. \textbf{14}, 282 (1980).
\bibitem{Schmit} E. Bogomolny and C. Schmit, \textit{Spectral statistics of a quantum interval-exchange map}, Phys. Rev. Lett. \textbf{93}, 254102 (2004).
\bibitem{Remy} E. Bogomolny, R. Dubertrand, and C. Schmit, \textit{Spectral statistics of a quantum interval-exchange map: the general case}, Nonlinearity, {\bf 22},  2101 (2009). 
\bibitem{GerlandEpjb} E.~Bogomolny, U.~Gerland, and C.~Schmit, \textit{Short-range plasma model for intermediate spectral statistics}, Eur. Phys. J. B {\bf 19}, 121 (2001).
\bibitem{ruijsenaars} S.N. Ruijsenaars, \textit{Complete Integrability of relativistic Calogero-Moser systems and elliptic function identities}, Commun. Math. Phys. \textbf{110}, 191 (1987).
 \bibitem{spin} J. Gibbons and T. Hermsen, \textit{A generalization of the Calogero-Moser system}, Physica \textbf{11D}, 337 (1984). 
\end{thebibliography}
\end{document}